\documentclass[acmsmall, screen]{acmart}

\AtBeginDocument{%
  }

\setcopyright{acmlicensed}
\copyrightyear{2018}
\acmYear{2018}
\acmDOI{XXXXXXX.XXXXXXX}

\acmJournal{JACM}
\acmVolume{37}
\acmNumber{4}
\acmArticle{111}
\acmMonth{8}

\usepackage[tt=false]{libertine}

\usepackage{url}
\usepackage{fancyvrb}
\usepackage[inline]{enumitem}  % Miscellaneous list-related commands
\usepackage{xspace}   % Smart spacing
\usepackage{supertabular}
\usepackage{bussproofs}

\usepackage{todonotes}

\usepackage{amsmath}
\usepackage{amsthm}

\usepackage{bm}       % Bold symbols in maths mode

\usepackage{mleftright} % Better \left and \right
\usepackage{pifont}

\usepackage{mathpartir}

\usepackage{thmtools, thm-restate}

\usepackage{mathtools}  % For "::=" ( \Coloneqq )

\usepackage{booktabs}   %% For formal tables:
\usepackage{subcaption} %% For complex figures with subfigures/subcaptions
\usepackage{longtable}  %% For spanning tables across multiple pages

\usepackage{listings}

\usepackage[lineBreakHack]{ottalt}

\usepackage{comment}

\usepackage{url}
\usepackage{
  nameref,%\nameref
  hyperref,%\autoref
}
\usepackage{tabularx}

\usepackage[flushleft]{threeparttable}
\newcommand{\concatOp}{+\kern-1.3ex+\kern0.8ex}  % http://tex.stackexchange.com/a/4195/73122

\theoremstyle{remark}
\newtheorem{theorem}{Theorem}[section]

\newtheorem{lemma}[theorem]{Lemma}

\newcommand{\typcast}[3]{#1 \hookrightarrow #2 : #3}

\newcommand{\mut}[2]{\mu #1.~#2}
\newcommand{\mat}[1]{\mut{\alpha}{#1}}
\newcommand{\tint}{\texttt{Int}}
\newcommand{\ttop}{\top}

\newcommand{\cid}{\textsf{id}}
\newcommand{\cfold}[1]{\textsf{fold}_{#1}}
\newcommand{\cunfold}[1]{\textsf{unfold}_{#1}}
\newcommand{\cseq}[2]{#1 ; #2}
\newcommand{\cfix}[2]{\textsf{fix}\,{#1}.~#2}

\newcommand{\folde}[2]{\textsf{fold}\,[#1]\,#2}
\newcommand{\unfolde}[2]{\textsf{unfold}\,[#1]\,#2}
\newcommand{\caste}[2]{\textsf{cast}\,[#1]\,#2}

\newcommand{\iscontract}[2]{#1 \text{ is contractive in } #2}

\newcommand{\name}{$\lambda^{\mu}_{Fi}$\xspace}
\newcommand{\nameS}{$\lambda^{\mu<:}_{Fi}$\xspace}
\newcommand{\graybox}[1]{\colorbox{gray!20}{#1}}
\newcommand{\grayboxm}[1]{\text{ \graybox{$ #1 $}}} % graybox in math mode

\newenvironment{ottdefnblock}[3][]{ \framebox{\mbox{#2}} \quad #3 \\[0pt]}{}

\newcommand{\ottsym}[1]{#1}

\newcommand{\ottafterlastrule}{\\}

  \renewottcommands[ott]

\begin{document}

\title{Full Iso-recursive Types}

\author{Litao Zhou}
\email{ltzhou@cs.hku.hk}
\orcid{1234-5678-9012}
\affiliation{%
  \institution{The University of Hong Kong}
  \streetaddress{Pokfulam Road}
  \city{Hong Kong}
  \country{China}
}

\author{Qianyong Wan}
\orcid{0009-0005-9894-2462}
\affiliation{%
  \institution{The University of Hong Kong}
  \streetaddress{Pokfulam Road}
  \city{Hong Kong}
  \country{China}
}
\email{qywan@cs.hku.hk}

\author{Bruno C. d. S. Oliveira}
\orcid{0000-0002-1846-7210}
\affiliation{%
  \institution{The University of Hong Kong}
  \streetaddress{Pokfulam Road}
  \city{Hong Kong}
  \country{China}}
\email{bruno@cs.hku.hk}


\begin{abstract}
There are two well-known formulations of recursive types: \emph{iso-recursive} and   
\emph{equi-recursive} types.
\citet{abadi1996syntactic} have shown that
iso- and equi-recursive types have the same expressive power.
However, their encoding of equi-recursive types in terms of
iso-recursive types requires explicit coercions. These coercions come
with significant
additional \emph{computational overhead}, and complicate reasoning
about the equivalence of the two formulations of recursive types.

This paper proposes a
generalization of iso-recursive types called \emph{full}
iso-recursive types. Full iso-recursive types allow encoding
all programs with equi-recursive types without
computational overhead. Instead of explicit term coercions, all 
type transformations are captured by \emph{computationally
irrelevant} casts, which can be erased at runtime without
affecting the semantics of the program. Consequently, reasoning about
the equivalence between the two approaches can be greatly simplified.
We present a calculus called \name, which extends the simply typed
lambda calculus (STLC) with full iso-recursive types. The \name
calculus is proved to be type sound, and shown to have the same
expressive power as a calculus with equi-recursive types.
We also extend our results to subtyping, and 
show that equi-recursive subtyping can
be expressed in terms of iso-recursive subtyping with cast operators.
\end{abstract}

\begin{CCSXML}
  <ccs2012>
    <concept>
        <concept_id>10003752.10003790.10011740</concept_id>
        <concept_desc>Theory of computation~Type theory</concept_desc>
        <concept_significance>500</concept_significance>
        </concept>
    <concept>
        <concept_id>10011007.10011006.10011008.10011009.10011011</concept_id>
        <concept_desc>Software and its engineering~Object oriented languages</concept_desc>
        <concept_significance>500</concept_significance>
        </concept>
  </ccs2012>
\end{CCSXML}

\ccsdesc[500]{Theory of computation~Type theory}
\ccsdesc[500]{Software and its engineering~Object oriented languages}

\keywords{Recursive types, Subtyping, Type system}

\received{20 February 2007}
\received[revised]{12 March 2009}
\received[accepted]{5 June 2009}

\maketitle

%%%%%%%%%%%%%%%%%%%%%%%%%%%%%%%%%%%%%%%%%%%%%%%%%%%%%%%%%%%%%%
% Section 1 Introduction
%%%%%%%%%%%%%%%%%%%%%%%%%%%%%%%%%%%%%%%%%%%%%%%%%%%%%%%%%%%%%%

\section{Introduction}
\label{sec:intro}
% Sketch for the argumentation of the paper:

% \begin{enumerate}

% \item There are two well-know formulations of recursive types: iso and  equi.

Recursive types are used in many programming languages to express
  recursive data structures, or recursive interfaces.
There are two well-known formulations of recursive types: \emph{iso-recursive} and   
  \emph{equi-recursive} types.
With equi-recursive types~\cite{morris1968lambda}, a recursive type $\mat{A}$ and its unfolding 
  $A[\alpha \mapsto \mat{A}]$ are equal, since they
  represent the same infinite tree~\cite{amadio1993subtyping}.
With iso-recursive types, a recursive type is 
only isomorphic to its unfolding~\cite{crary1999recursive}.
To witness the isomorphism, explicit fold and unfold operators are used.

Because both formulations provide alternative ways to model recursive
types, the relationship between iso- and equi-recursive types
  has been a topic of
  study~\cite{abadi1996syntactic,patrignani2021semantic,urzyczyn1995positive}.
  Understanding this relationship is important to
  answer questions such as whether the expressive power of the two
  formulations is the same or not.
\citeauthor{urzyczyn1995positive} proved that these two 
  formulations have the same expressive power
  when the types considered are restricted to be positive.
  \citeauthor{abadi1996syntactic} extended
  \citeauthor{urzyczyn1995positive}'s result and showed that
  unrestricted formulations of iso- and equi-recursive types also have
  the same expressive power, leading to the well-known statement that 
  ``iso-recursive types have the same expressive power as 
  equi-recursive types''. In addition, 
\citeauthor{patrignani2021semantic} 
  showed that the translation from iso-recursive to equi-recursive types
  is fully abstract with respect to contextual equivalence.
%  Our work aligns more closely with 

  However, the encoding proposed by \citeauthor{abadi1996syntactic} requires explicit coercions, which are
  interpreted as functions to be evaluated at runtime.
  %This results in
  %significant \emph{computational overhead}. Thus iso-recursive
  Iso-recursive types can only encode equi-recursive types with significant
  additional \emph{computational overhead}.
  Moreover, these explicit
  coercions cannot be easily erased and therefore complicate the
  reasoning about \emph{behavioral equivalence}.  To address the latter
  challenge, \citeauthor{abadi1996syntactic} defined an axiomatized
  program logic and showed that the iso-recursive term obtained by
  their encoding behaves in the same way as the original equi-recursive term
  in the logic. However, the soundness of their program logic is left
  as a conjecture, since they did not consider an operational semantics
  in their work. Thus, behavioral equivalence between programs
  written with equi-recursive and iso-recursive types lacks a complete
  proof in the literature. Without introducing explicit coercions, iso-recursive types are
  strictly weaker than equi-recursive types, since the infinite tree
  view of equi-recursive types equates more types than isomorphic
  unfoldings of recursive types.

  This paper proposes a
  \emph{generalization} of iso-recursive types called \emph{full}
  iso-recursive types.
  Full iso-recursive types overcome the challenges of traditional iso-recursive types in
  achieving the typing expressiveness and behavioral equivalence seen
  in equi-recursive types. Instead of fold and
  unfold operators and explicit coercions, we use a more general
  notion of \emph{computationally irrelevant cast
    operators}~\cite{sulzmann2007system, cretin2014erasable}, which allow transformations on any types that are
  equivalent in an equi-recursive setting. Full iso-recursive types can encode
  \emph{all} programs with equi-recursive types \emph{without}
  computational overhead, since casts can be erased at runtime without
  affecting the semantics of the program. Consequently, 
  the semantic equivalence between programs written with
  equi-recursive and full iso-recursive types is also greatly
  simplified, and allows for a complete proof, compared to
  \citeauthor{abadi1996syntactic}'s work.

  We present a calculus called \name, which extends the simply typed
  lambda calculus (STLC) with full iso-recursive types. The \name
  calculus is proved to be type sound, and shown to have the same
  typing power as a calculus with equi-recursive types.
%  To show that full iso-recursive
%  types have the same typing power as equi-recursive types,
  To prove the latter result, we 
  define a type-directed elaboration from the calculus with equi-recursive types to
  \name, and an erasure function that removes all
  casts from full iso-recursive terms to obtain equi-recursive terms.
  Moreover, the termination and divergence behavior of programs is
  preserved under the elaboration and erasure operations.  Therefore,
  \name is sound and complete w.r.t. the calculus with equi-recursive types in terms of
  both typing and dynamic semantics.  On the other hand, traditional
  iso-recursive types can be seen as a special case of full
  iso-recursive types. One can easily recover the traditional
  unfold and fold operators by using the corresponding cast operators accordingly.
  So all the results for iso-recursive types can be adapted to full
  iso-recursive types as well.

% \item We also extend our results to subtyping and show that equi-recursive subtyping can
%   be expressed in terms of iso-recursive subtyping with full casts.
%   Previous work has not covered subtyping. (EXPAND ON THIS) 
We also extend our results to subtyping and 
  show that equi-recursive subtyping can
  be expressed in terms of iso-recursive subtyping with cast operators.
Although subtyping between equi-recursive types~\cite{brandt1998coinductive,amadio1993subtyping,gapeyev2002recursive} and
  subtyping between iso-recursive types~\cite{abadi2012theory,zhou2022revisiting} has been studied
  in depth respectively in the literature,
  the relationship between the two approaches has been largely unexplored. 
%  probably because the relation between equi-recursive and iso-recursive
%  types without subtyping is already complex, as discussed above.
We revisit \citet{amadio1993subtyping}'s seminal work on
  equi-recursive subtyping and observe that an equi-recursive
  subtyping relation can be decomposed into a combination of
  equi-recursive equalities and an iso-recursive subtyping relation.
Since our cast operators can capture all the equi-recursive equalities,
  we can achieve a simple encoding of equi-recursive subtyping
  in the setting of full iso-recursive types with subtyping.

% \item Summary of key technical results: what do we show? Maybe create a Diagram/Figure.
% maybe this has been covered by the contributions part

% \item 
Full iso-recursive types open the path for new applications.
  For example, in the design of realistic compilers,
  it is common to have source languages that are lightweight in terms of type annotations;
  and target languages, which are used internally, that are heavy on annotations, but are simple to type-check.
  For instance, the GHC Haskell compiler works in this way: the source language (Haskell) has a lot of convenience
  via type inference, and no explicit casts are needed in source programs. A source program is then elaborated
  to a variant of System Fc~\cite{sulzmann2007system}, which is a System F like language with explicit type annotations, type applications and
  also explicit casts. Our work enables designing source languages with equi-recursive types, which
  are elaborated to target languages with full iso-recursive types.
  Equi-recursive types offer convenience because they can avoid explicit folds and unfolds,
  but type-checking is complex. With full iso-recursive types we need to write explicit casts,
  but type-checking is simple. Thus we can have an architecture similar to that of GHC.  
  In this scenario it is important that no computational overhead is introduced during the elaboration,
  which is why using standard iso-recursive types would not be practical.
  In addition, source languages could also hide explicit casts into language constructs (such as constructors, method calls
  and/or pattern matching). This would be another way to use full iso-recursive types, which is similar to current applications
  of iso-recursive types.

% \end{enumerate}

The main contributions of this paper are as follows:
\begin{itemize}
\item \textbf{Full iso-recursive types.} 
  We propose a novel formulation of recursive types, called full iso-recursive types, 
  which generalizes the traditional iso-recursive fold and unfold operators to 
  cast operators.
\item \textbf{The \name calculus.}
  We introduce the \name calculus, which extends the simply typed lambda calculus 
  with full iso-recursive types.
  The calculus is equipped with a type system, a call-by-value operational semantics, 
  and a type soundness proof.
\item \textbf{Equivalence to equi-recursive types.}
  We show that \name is equivalent to STLC extended with
  equi-recursive types in terms of typing and dynamic semantics.
\item \textbf{Extension to subtyping.}
  We present \nameS, an extension of \name with iso-recursive subtyping,
  and show the same metatheory results for \nameS,
  namely, type soundness, typing equivalence and behavioral equivalence to
  equi-recursive types with subtyping.
\item \textbf{Coq formalization.}
  We provide a mechanical formalization and proofs for 
  all the new metatheory results of full iso-recursive types in Coq,
  except for Theorem~\ref{thm:equi-to-iso}, which is adapted from the literature~\cite{amadio1993subtyping}.
\end{itemize}

%%%%%%%%%%%%%%%%%%%%%%%%%%%%%%%%%%%%%%%%%%%%%%%%%%%%%%%%%%%%%%
% Section 2 Overview
%%%%%%%%%%%%%%%%%%%%%%%%%%%%%%%%%%%%%%%%%%%%%%%%%%%%%%%%%%%%%%

\section{Overview}
This section provides an overview of our work. We first briefly
review the two main approaches to recursive types, namely
iso-recursive types and equi-recursive types, and the relationship
between the two approaches. Then we introduce our key ideas and results.

\subsection{Equi-recursive Types}

Equi-recursive types treat recursive types and their unfoldings as equal.
% In \rref{Typing-eq}, a special relation $\tyeq{A}{B}$ is defined to 
%   compare two equi-recursive types for equality.
The advantage of equi-recursive types is that they are simple to use,
  since there is no need to insert explicit annotations in the term language
  to transform between equal types, as shown in \rref{Typ-eq}.
$$
\drule{Typ-eq}
$$
The metatheory of equi-recursive types has been comprehensively studied by 
  \citet{amadio1993subtyping}.
They proposed a tree model for specifying equality (or subtyping) between
  equi-recursive types.
In essence, two recursive types are equal (or subtypes) if their infinite
  unfoldings are equal (or in a subtyping relation).
The tree model provides a clear and solid foundation for the interpretation 
  of equi-recursive types.

\begin{figure}
  \drules[Tyeq]{$A\doteq B$}{Equi-recursive Equality}
    {contract, unfold, mu-cong, trans, refl, symm, arr}
  \caption{\citeauthor{amadio1993subtyping}'s equi-recursive type equality.}
  \label{fig:ac-tyeq}
\end{figure}

\citeauthor{amadio1993subtyping} also provided a rule-based axiomatization 
  to compare equi-recursive types, as shown in Figure~\ref{fig:ac-tyeq}. They proved the soundness and completeness
  of the rules to the tree-based interpretation.
For example, \rref{Tyeq-unfold} states that a recursive type is equal
  to its unfolding, and \rref{Tyeq-mu-cong} states that the equality
  is congruent with respect to the recursive type operator.
% the contract rule has a reference to "A semantic basis for Quest" in the Amadio & Cardelli paper
\Rref{Tyeq-contract} states that two types are equal if
  they are the fixpoints of the same type function $A[\alpha]$.
Note that $A$ needs to be contractive in $\alpha$, i.e. either
  $\alpha$ is not free in $A$ or $A$ can be unfolded to a type
  of the form $A_1 \rightarrow A_2$.
This is to prevent equating arbitrary types using non-contractive
  type functions, such as when $A$ is $\alpha$.
\Rref{Tyeq-contract} allows recursive types that have equal infinite
  unfoldings, but are not directly related by finite unfoldings, to be equal.
For example, let $A[\alpha] = \tint \to \tint \to \alpha$, 
  then $B_1 = \mu\alpha.\tint \to \alpha$ and $B_2 = \mu\alpha.\tint \to \tint \to \alpha$
  are equal according to \rref{Tyeq-contract}:
\begin{center}
  \begin{prooftree}
    \AxiomC{\ldots}
    \UnaryInfC{$B_1 \doteq A[\alpha\mapsto B_1]$}
    \AxiomC{}
    \RightLabel{\rref*{Tyeq-unfold}}
    \UnaryInfC{$B_2 \doteq A[\alpha\mapsto B_2]$}
    \AxiomC{$\iscontract{A}{\alpha}$}
    \RightLabel{\rref*{Tyeq-contract}}
    \TrinaryInfC{$\mu\alpha.\tint \to \alpha \doteq \mu\alpha.\tint \to \tint \to \alpha$}
  \end{prooftree}
\end{center}

\noindent Here, the missing derivation is:

\begin{center}
  \begin{minipage}[H]{0.83\textwidth}
    \begin{prooftree}
      \AxiomC{}
      \RightLabel{\rref*{Tyeq-refl}}
      \UnaryInfC{$\tint \doteq \tint$}
      \AxiomC{}
      \RightLabel{\rref*{Tyeq-unfold}}
      \UnaryInfC{$\mat{\tint\to\alpha} \doteq \tint \to \mat{\tint\to\alpha}$}
      \RightLabel{\rref*{Tyeq-arrow}}
      \BinaryInfC{$\tint\to B_1 \doteq \tint\to\tint\to B_1$}
      \RightLabel{\rref*{Tyeq-trans,Tyeq-unfold}}
      \UnaryInfC{$B_1 \doteq A[\alpha\mapsto B_1]$}
    \end{prooftree}  
  \end{minipage}
\end{center}

Despite its equivalence to the tree model, \citeauthor{amadio1993subtyping}'s 
  axiomatization is not easy to use in practice. In particular one needs
  to find a generating type function $A[\alpha]$ in \rref{Tyeq-contract}.
Later on, there have been a few alternative axiomatizations of 
  equi-recursive types~\cite{brandt1998coinductive,danielsson2010subtyping,gapeyev2002recursive},
  which are all proved to be equivalent to the tree model.
Among them, \citeauthor{brandt1998coinductive} proposed an inductively defined relation
  $H \vdash A \doteq B$ for equi-recursive type equality, shown in Figure \ref{fig:equi-equal}.
$H$ is a list of type equality assumptions that can be used to derive the
  equality $A \doteq B$.
New equalities are added to $H$ every time function types are compared, as shown
  in \rref{Tye-arrfix}.
Compared to \rref{Tyeq-contract}, \rref{Tye-arrfix} encodes the coinductive
  essence of equi-recursive types in a simpler way.
Therefore, we choose \citeauthor{brandt1998coinductive}'s axiomatization
  as the basis for our work.

\begin{figure}
  \centering
  \drules[Tye]{$H \vdash A \doteq B$}{Inductive Equi-recursive Equality}
    {assump, refl, trans, unfold, symm, arrfix}
  \caption{\citeauthor{brandt1998coinductive}'s inductively defined equi-recursive type equality.}
  \label{fig:equi-equal}
\end{figure}

\subsection{Iso-recursive Types}

Iso-recursive types~\cite{crary1999recursive} are a different approach that treats recursive types
  and their unfoldings as different, but isomorphic up to an unfold/fold
  operator. With iso-recursive types foldings and unfoldings of the
  recursive types must be explicitly triggered, and there is no typing
  \rref{Typ-eq} to implicitly convert between equivalent types.
\Rref{Typ-unfold} and \rref{Typ-fold} show the typing rules for unfolding and
  folding a term of recursive types.
A \textsf{fold} expression constructs a recursive type, while an \textsf{unfold} 
  expression opens a recursive type to its unfolding.

\begin{center}
  \text{\drule[]{Typ-unfold}}
  \quad\quad\quad
  \text{\drule[]{Typ-fold}}
\end{center}
%Without the type equality rules, iso-recursive types simplify
%typing.
One advantage of iso-recursive types is that
they are easier to extend to more complex type systems, which may
easily make the type equality relation undecidable. Instead, iso-recursive
types provide explicit control over folding and unfolding, 
avoiding issues with undecidability.
One disadvantage of iso-recursive
  types is their inconvenience in use due to the explicit fold and 
  unfold operators. However, this disadvantage can be mitigated by hiding folding and unfolding under
  other language constructs, such as pattern matching,
  constructors or method calls~\cite{crary1999recursive, lee2015theory,
    zhou2022revisiting, pierce2002types, harper2000type, vanderwaart2003typed,
    yang2019pure}.
As we shall see in Section~\ref{subsec:relating}, a further disadvantage of
iso-recursive types is that folding and unfolding alone is not enough
to provide all of the expressive power of the type equality rules. In
some cases, explicit, computationally relevant, term coercions are necessary.

\subsection{Relating Iso-recursive and Equi-recursive Types}\label{subsec:relating}

The relationship between iso-recursive types and equi-recursive types has been a
subject of study for a long time on the literature of recursive
types~\cite{abadi1996syntactic,patrignani2021semantic, urzyczyn1995positive}.
This subsection reviews the existing approaches to relate the two approaches
  and their issues.

\paragraph{Encoding iso-recursive types.}
The encoding of iso-recursive types in equi-recursive types is straightforward, simply
  by erasing the fold and unfold operators~\cite{abadi1996syntactic}.
Since the \rref{Tyeq-unfold} states that a recursive type is equal to its unfolding,
  it is easy to see that the encoding is type preserving.
The encoding is also behavior preserving, since the reduction rules with fold and unfold
  operators will become no-ops when erased, as shown below:
\begin{center}
  \text{\drule[]{Red-fld}}
  \quad
  \text{\drule[]{Red-ufd}}
  \quad
  \text{\drule[]{Red-elim}}
\end{center}

Notice that in the process of reducing folded and unfolded
expression $e$, we merely reduce $e$. The type $A$ does not
influence the reduction of $e$. Eventually, when $e$ reaches a value $v$,
an unfold cancels a fold and we simply obtain $v$.
In other words, folding and unfolding are \emph{computationally
  irrelevant}: they do not influence the runtime result, and can be
erased, to avoid runtime costs. 
Moreover, \citet{patrignani2021semantic} proved that the
  erasure operation is fully abstract, i.e. two terms that cannot be
  distinguished by any program contexts in the iso-recursive setting 
  are also indistinguishable in the equi-recursive setting.
% This is a stronger result than semantic equivalence for closed terms.

\paragraph{Encoding equi-recursive types via fold and unfold}
It takes more effort to encode equi-recursive types in terms of iso-recursive types.
Since equi-recursive types treat recursive types and their unfoldings as equal,
  we need to insert explicit fold and unfold operators in the iso-recursive setting
  to transform between equal types.
For example, let $e$ be a function that keeps taking integer arguments and
  returning itself, which can be typed as a recursive type $\mat{\tint \to \alpha}$.
In an equi-recursive setting, $(e\,1)$ can be typed as $\mat{\tint \to \alpha}$,
  by using the \rref{Typ-eq} and \rref{Tyeq-unfold} to unfold the recursive type
  to $\tint \to (\mat{\tint \to \alpha})$ so that it can be applied to the argument $1$.
However, in the iso-recursive setting, we need to insert an unfold operator
  to make the transformation explicit, as shown in the following derivation:
\begin{center}
  \begin{prooftree}
      \AxiomC{$\vdash e : \mat{\tint\to\alpha}$}
      \LeftLabel{\rref*{Typ-unfold}}
      \UnaryInfC{$\vdash \unfolde{\mat{\tint \to \alpha}}{e} : \tint \to (\mat{\tint\to\alpha}) $}
      \AxiomC{$\vdash 1 : \tint$}
      \LeftLabel{\rref*{Typ-app}}
      \BinaryInfC{$\vdash (\unfolde{\mat{\tint \to \alpha}}{e}) \, 1 : \mat{\tint\to\alpha} $}
  \end{prooftree}
\end{center}

\paragraph{Fold/unfold is not enough: computationally relevant explicit coercions}
The above example shows that, for some equi-recursive terms, inserting fold and unfold 
  operators within the term language can achieve an encoding in terms
  of iso-recursive types.
However, this is not always the case.
Recall that $\mat{\tint \to \alpha}$ and $\mat{\tint \to \tint \to \alpha}$ are also equal
  in the equi-recursive setting, but they are not directly related by fold and unfold operators.
To address this issue, \citet{abadi1996syntactic} proposed an approach to insert
  \emph{explicit coercion functions}.
They showed that, for any two equi-recursive
  types $A$ and $B$ considered to be equal following the derivation in Figure~\ref{fig:ac-tyeq}, 
  there exists a coercion function $f: A \to B$
  that can be applied to terms of type $A$ to obtain terms of type $B$.
With the coercion function, terms that are well typed by \rref{Typ-tyeq} can now have an
  encoding in terms of iso-recursive types, possibly with the help of explicit coercion functions.

One issue is that the insertion of coercion functions affects the computational
  structure of the terms.
For example, assume that $e$ has a function type $\tint \to \tint \to (\mat{\tint \to \alpha})$.
This type can be partially folded to $\tint \to (\mat{\tint \to
  \alpha})$. In an equi-recursive setting, due to the \rref{Typ-eq}, the term $e$ can also be assigned the type $\tint \to (\mat{\tint \to
  \alpha})$, without any changes. 
In an iso-recursive setting, in addition to folding and unfolding, we
need explicit coercions.
The coercion function for this transformation is 
$$
\lambda(x: \tint \to \tint \to \mat{\tint \to \alpha}).~\lambda(y: \tint).~
  \folde{\mat{\tint \to \alpha}}{ (x\, y) }
$$
Now, applying the coercion function to $e$ results in a term of type $\tint \to (\mat{\tint \to \alpha})$.
Unfortunately, such explicit coercion functions are computationally
relevant. Thus, an encoding of equi-recursive types in terms of
iso-recursive types can introduce non-trivial computational overhead.
The issue is particularly problematic because some coercions need to
essentially be \emph{recursive} functions. Therefore, it is impractical to
use such an encoding in a language implementation.

\paragraph{Issues with reasoning} Explicit coercions also bring
  new challenges in terms of reasoning, and in particular in proving
  the behavioral preservation of the encoding.
Continuing with the previous example, if we transform this resulting term back to an equi-recursive setting, by
  erasing the fold and unfold operators, we will get a term:
\begin{equation}
  (\lambda(x: \tint \to \tint \to \mat{\tint \to \alpha}).~\lambda(y: \tint).~ (x\, y)) \, e  
  \label{eqn:encode-example}
\end{equation}
This term is equivalent to $e$ under $\beta-$ and $\eta-$reduction, but it is not the
  same as $e$ anymore.
In more complicated cases, especially for derivations involving the use of \rref{Tyeq-contract},
  the insertion of coercion functions can lead to a significant
  change in the syntactic structure of the terms, which makes it difficult to reason
  about the behavior preservation of the encoding. In essence, with the
  \rref{Tyeq-contract}, one needs
  to encode special iterating functions to model the fixpoint of a type function.
\citeauthor{abadi1996syntactic} proved that the encoding is equivalent 
  to the original term in an axiomatized program logic, 
  but the soundness of the program logic is conjectured
  to be sound, and the authors did not consider an operational
  semantics. Thus, while it is expected that the behavioral
  equivalence result holds (assuming the conjecture and a suitable
  operational semantics), there is no complete proof in the
  literature for this result. 

\subsection{Subtyping}
\label{subsec:subtyping}
\paragraph{Equi-recursive subtyping}
It is common to extend recursive types with subtyping.
For equi-recursive types, \citeauthor{amadio1993subtyping} proposed a
set of rules, which relies on the equality relation in Figure~\ref{fig:ac-tyeq}.
We show some selected rules below:
\begin{displaymath}
  \text{ \drule[]{ACSub-eq} }
  \quad
  \text{ \drule[]{ACSub-arrow} }
  \quad
  \text{ \drule[]{ACSub-rec} }
  \quad
  \text{ \drule[]{ACSub-var} }
\end{displaymath}
Two types are in a subtyping relation if their infinite unfoldings are equal,
  as shown in \rref{ACSub-eq}.
The subtyping relation is structural, as can be seen in \rref{ACSub-arrow}.
For dealing with recursive types, \rref{ACSub-rec} states that two recursive types
  are in a subtyping relation if their recursive bodies are subtypes,
  when assuming that the recursive variable of the two types are in a subtyping relation.
The subtyping rules are also referred to as the Amber rules, since  \rref{ACSub-rec}
  is adopted by the implementation of the Amber programming 
  language~\cite{cardelli1985amber}.
The Amber rules are proved to be sound and complete to the tree model interpretation
  of equi-recursive subtyping~\cite{amadio1993subtyping}.

\paragraph{Iso-recursive subtyping}
For iso-recursive types, one can replace the equi-recursive equality relation 
  in \rref{ACSub-eq} with the syntactic equality relation to obtain the
  iso-recursive style Amber rules.
The iso-recursive Amber rules are well-known and widely used for subtyping 
  iso-recursive types~\cite{abadi2012theory,bengtson2011refinement,chugh2015isolate,
    lee2015theory, swamy2011secure, duggan2002type}. 
However, the metatheory for the iso-recursive Amber rules has not been well studied
  until recently~\cite{zhou2022revisiting, zhou2020revisiting}.
\citeauthor{zhou2022revisiting} provided a new specification for 
  iso-recursive subtyping and proved a number of metatheory results,
  including type soundness, transitivity of the subtyping relation, and
  equivalence to the iso-recursive Amber rules.

However, unlike type equality, the relation between equi-recursive and 
  iso-recursive subtyping has been less studied.
One attempt that we are aware of is the work by \citet{ligatti2017subtyping}.
They provided an extension of the iso-recursive
  subtyping rules to allow for subtyping between recursive types and
  their unfoldings, but their rules cannot account for the full
  expressiveness of equi-recursive subtyping.
For example, $\mat{\tint\to\alpha} \le \mat{\tint\to\tint\to\alpha}$
  is a valid subtyping relation in the equi-recursive Amber rules using
  \rref{ACSub-eq}, but it is not derivable in \citeauthor{ligatti2017subtyping}'s
  rules.

\subsection{Key Ideas and Results}
\label{subsec:results}

As we have shown, encoding iso-recursive types with equi-recursive types is simple.
As for the other direction, \citeauthor{abadi1996syntactic} showed that iso-recursive types can be
  encoded with equi-recursive types, which leads to a well-known statement
  that ``iso-recursive types have the same expressive power as equi-recursive types''.
However, their encoding involves the insertion of explicit coercion functions,
  and lacks a complete proof of correctness.
In our work, we present a novel approach to iso-recursive types, 
  full iso-recursive types, which extends the unfold and fold operators to
  a more general form. We show that full iso-recursive types and
  equi-recursive types can be mutually encoded and
  the encoding preserves the semantic behavior.
Compared to the previous work, the correctness proof of our encoding is 
  straightforward and foundational,
  without relying on any a~priori assumptions.

\paragraph{Type Casting.}
The key idea of our approach is the introduction of a type casting relation
  that generalizes the unfold and fold operators.
Instead of allowing only the unfold and fold operators to transform between
  recursive types and their unfoldings, we allow terms of any type
  to be transformed to their equivalent type using the type casting relation.
The \rref{Typ-unfold,Typ-fold} are now replaced by the following rule:
\begin{center}
  \drule{Typ-cast}
\end{center}
The type casting relation $\typcast{A}{B}{c}$ states that type $A$ can be 
  cast to type $B$ using the casting operator $c$.
Essentially, the type casting relation is an equivalent form of
  \citeauthor{brandt1998coinductive}'s type equality relation, augmented
  with a casting operator $c$, a new syntactic construct to witness the
  proof of the type casting relation.
As we will show in the following sections, the type casting relation
  is not a simple extension of Figure~\ref{fig:equi-equal}.
For example, we remove the \rref{Tye-symm} from the type casting relation,
  which is hard to interpret in the operational semantics, and proved that
  it is admissible from the remaining rules.
After resolving these issues, it is easy to encode the equi-recursive
  \rref{Typ-eq} using the type casting relation in \rref{Typ-cast}.
For instance, the encoding in (\ref{eqn:encode-example}) can now
  be replaced by the following term:
  $$
  \caste{\cid \to \cfold{(\mat{\tint \to \alpha})} }{e}
  $$
Here $\cid$ is the identity casting operator, and $\cfold{\mat{\tint \to \alpha}}$
  is the casting operator that witnesses the proof of a folding from
  $\tint \to (\mat{\tint \to \alpha})$ to $\mat{\tint \to \alpha}$.
Thus, the full iso-recursive typing rules are equivalent to the equi-recursive
  typing rules.

On the other hand, the unfolding and folding operators in standard
  iso-recursive types can be recovered from our type casting relation.
For example, the term $(\caste{\cfold{A}}{e})$ is essentially equivalent to
  $(\folde{A}{e})$, and the term $(\caste{\cunfold{A}}{e})$ is
  equivalent to $(\unfolde{A}{e})$ in terms of typing and dynamic
  semantics, as we will show in the following sections.
Therefore, full iso-recursive types are a generalization of
standard iso-recursive types.

\paragraph{Push Rules.}
The extension of the typing rules brings new challenges to the
  design of semantics and the proof of type soundness.
With the casting operator, there can be terms that are not simple
  unfoldings or foldings of recursive types, and the operational
  semantics needs to be extended to handle these terms.
For example, terms such as $(\caste{\cid \to \cfold{(\mat{\tint \to
      \alpha)}} }{e})$, which have no analogous representation in
calculi with standard iso-recursive types, need to be considered
during reduction.
To address this issue, we introduce a set of new reduction rules
  to handle casting operators:
\begin{displaymath}
  \begin{array}{c}
    \text{\drule[]{Red-cast}}
    \quad
    \text{\drule[]{Red-cast-id}} \\
    \, \\
    \text{\drule[]{Red-cast-arr}} \\ \,
  \end{array}
\end{displaymath}
The reduction rules are designed in a call-by-value fashion, and
  we also define the cast of values of function types (i.e. $\caste{c_1\to c_2}{v}$)
  as values.
The new reduction rules in our system are referred to as \emph{push} rules,
  since they push the casting operators inside the terms to make
  the terms more reducible, as shown in \rref{Red-Cast-Arr}.
Our design is inspired by the homonymous push rules in the design of
  calculi with coercions~\cite{cretin2014erasable, sulzmann2007system}.
Note that the casting operator $\neg c_1$ computes the dual of the casting operator $c_1$,
  which is used to indicate the reverse transformation that $c_1$ represents.
This operation is necessary to ensure that the reduction is type preserving,
  since applying a casting operator to the expected input type of a function
  is essentially equivalent to applying the reverse 
  casting to the actual input argument.
A running example of a reduction using the push rules is shown as follows:
  $$
  \begin{array}{rll}
  & (\caste{\cid \to \cfold{(\mat{\tint \to \alpha})} }{v})\,1  \\
  \hookrightarrow & \caste{\cfold{(\mat{\tint \to \alpha})}}{(v\,(\caste{\cid}{1}))} & (\text{\rref*{Red-cast-arr}}) \\
  \hookrightarrow & \caste{\cfold{(\mat{\tint \to \alpha})}}{(v\,1)} & (\text{\rref*{Red-cast-id} and \rref*{Red-cast}})
  \end{array}
  $$

One of the key results of our work is the type soundness of the full iso-recursive
  calculus, which is proved by showing that the push rules preserve the type
  of the terms and the type casting relation.
This is one step beyond \citeauthor{brandt1998coinductive}'s work,
  in which a coercion typing rule similar to our casting rules 
  was introduced, but no results about the dynamic semantics were studied.
Another contribution of our work is that with full iso-recursive types,
  we retain the computational structure of the terms when encoding 
  equi-recursive types.
In other words, erasing the casting operators from the terms will result in
  the original terms, which is not the case for the previous
  work~\cite{abadi1996syntactic}.
  Our casts are \emph{computationally irrelevant}, and unlike regular
  iso-recursive types, which require computationally relevant term
  coercions for some type conversions, no such coercions are needed in our approach. 
For example, all the reduction steps in the example above
are erased to the original term $(v\,1)$.
This round-tripping property simplifies the correctness reasoning of the encoding.

We show that all the terms that are well-typed in the equi-recursive setting
  can be encoded in the full iso-recursive setting. Furthermore, the encoding is
  behavior preserving, i.e. evaluating the encoded terms will result in a
  value that is equal to the value of the original equi-recursive term
  up to erasure.
In this sense, we get back a fully verified statement that \emph{``full 
  iso-recursive types have the same expressive power as equi-recursive types''}.

\paragraph{Subtyping} Our results extend to subtyping.
Our main observation for subtyping is to show that 
  the equi-recursive subtyping relation can be
  defined by a combination of equi-recursive equality and
  the iso-recursive subtyping relation~\cite{cardelli1985amber,abadi2012theory, zhou2022revisiting}, 
  as shown below:
$$
A \le_e B \triangleq \exists\, C_1\, C_2.~  (A \doteq C_1) \land (C_1 \le_i C_2) \land (C_2 \doteq B)
$$
Here $\le_e$ denotes an equi-recursive subtyping relation, and $\le_i$
denotes an iso-recursive subtyping relation. This alternative
definition of equi-recursive subtyping is implicitly implied from
\citeauthor{amadio1993subtyping}'s work, but it is somewhat hidden
behind their proofs and definitions. We reinterpret their proofs and definitions to
highlight that this alternative definition
is equivalent to existing equi-recursive subtyping definitions in the
literature.

This alternative definition of equi-recursive subtyping is important
because we can reuse the existing type casting relation in the full
iso-recursive setting with subtyping.
For example, given an equi-recursive term $e$ that has the type $A$
  with $A \le_e B$, we can encode $e : B$ in the full iso-recursive setting
  as $((\caste{c_2}{(\caste{c_1}{e'})}) : B)$, in which $c_1$ and $c_2$ are casts
  encoding the equality relation $A \doteq C_1$ and $C_2 \doteq B$, and
  $e'$ is the encoding of $(e:A)$.
This term type checks with the iso-recursive subtyping relation.

% Moreover, since the casting operators remain erasable,
%   we also retain the semantic preservation proof of the encoding.
Our encoding is still computationally irrelevant
  in the presence of subtyping. Thus, all the results -- including type soundness, well-typed encoding, and
  behavior preservation -- are also applicable to the system
  with subtyping.
%  at the cost of only a few additional technical details.
This is a significant improvement over previous 
  work~\cite{abadi1996syntactic}, which has not studied the
  relationship between equi-recursive and iso-recursive subtyping.

%%%%%%%%%%%%%%%%%%%%%%%%%%%%%%%%%%%%%%%%%%%%%%%%%%%%%%%%%%%%%%
% Section 3 A calculus with full iso-recursive types
%%%%%%%%%%%%%%%%%%%%%%%%%%%%%%%%%%%%%%%%%%%%%%%%%%%%%%%%%%%%%%

\section{A calculus with full iso-recursive types}

\label{sec:calculus}

In this section we will introduce a calculus with full iso-recursive types,
  called \name.
Our calculus is based on the simply typed lambda calculus extended
  with recursive types and type cast operators.

\subsection{Syntax and Well-formedness}
\label{subsec:syntax}
The syntax of \name is shown at the top of Figure~\ref{fig:syntax}.

\begin{figure}[t]
\begin{tabular}{llll}%\toprule
  Types & $A, B$ & $\Coloneqq$ & $\tint %\mid \ttop   
                \mid A_1 \to A_2  \mid  \alpha \mid \mu \alpha . ~A$ \\
  Expressions & $e$ & $\Coloneqq$ & $x \mid \textsf{n}  \mid e_1~e_2 \mid \lambda x:A .~ e \mid \caste{c}{e}$ \\
  Values & $v$ & $\Coloneqq$ & $ \textsf{n}  \mid \lambda x : A .~ e \mid \caste{\cfold{A}}{v} \mid \caste{c_1 \to c_2}{v}$ \\
  Cast Operators & $c$ & $\Coloneqq$ & $ \iota \mid \cid \mid \cfold{A} \mid \cunfold{A} \mid c_1 \to c_2 \mid \cseq{c_1}{c_2} \mid \cfix{\iota}{c}$ \\
  Type Contexts & $\Delta$ & $\Coloneqq$ & $\cdot \mid  \Delta, \alpha $ \\
  Typing Contexts & $\Gamma$ & $\Coloneqq$ & $\cdot \mid  \Gamma, x:A  $ \\
  Type Cast Contexts & $\mathbb{E}$ & $\Coloneqq$ & $\cdot \mid  \mathbb{E}, \iota : A \hookrightarrow B  $ \\
\end{tabular}
\centering
  \drules[WFT]{$\Delta \vdash \textit{A}$}{Well-formed Type}{%top, 
      int, var, arrow, rec}
\caption{Syntax and type well-formedness of \name.}
\label{fig:syntax}
\end{figure}

\paragraph{Types.} Meta-variables $A, B$ range over types. 
Types include base types ($\tint$), % the top type ($\ttop$), 
  function types ($A_1 \to A_2$), type variables ($\alpha$), 
  and recursive types ($\mu \alpha . ~A$).

\paragraph{Expressions.} Meta-variables $e$ range over expressions.
Most of the expressions are standard, including: variables ($x$), integers ($\textsf{n}$), 
  applications ($e_1~e_2$) and lambda abstractions ($\lambda x:A .~ e$). 
We also have a type cast operator ($\caste{c}{e}$) that
  transforms the type of the expression $e$ to an equivalent type
  using the cast operator $c$.
The cast operators $c$ include 
  cast variables ($\iota$),
  the identity cast ($\cid$),
  the fold and unfold casts ($\cfold{A}$ and $\cunfold{A}$),
  the arrow cast ($c_1 \to c_2$), the sequential cast ($\cseq{c_1}{c_2}$),
  and the fixpoint cast ($\cfix{\iota}{c}$).
We will define the type cast rules for these operators in 
  \S\ref{subsec:typing}.

\paragraph{Values.} Meta-variables $v$ range over values.
Integers ($\textsf{n}$) and lambda abstractions ($\lambda x : A .~ e$),
  are considered as values, which are standard for a simply typed lambda calculus.
In standard iso-recursive types, the folding of a value ($\folde{A}{v}$)
  is a value.
Therefore in our calculus its corresponding encoding  ($\caste{\cfold{A}}{v}$)
  is also considered a value.
We also consider arrow casts of a value ($\caste{c_1 \to c_2}{v}$) to be values,
since they cannot be reduced further.

\paragraph{Contexts and Well-formedness}
Type contexts $\Delta$ track the bound type variables $\alpha$.
A type is well-formed if all of its free variables are in the context. 
The well-formedness rules for types are standard, and shown at the bottom Figure~\ref{fig:syntax}.
Typing contexts $\Gamma$ track the bound variables $x$ with their types.
A typing context is well-formed ($\vdash \Gamma$) if there are no duplicate variables
  and all the types are well-formed.
We also define a type cast context $\mathbb{E}$ to keep track of the
  cast variables $\iota$ and the cast operators that they are associated with.
This will be used in the type cast rules, which we will define in \S\ref{subsec:typing}.

For type variables and term variables, we assume the usual notions of 
  free and bound variables, and the usual capture-avoiding substitution 
  function, denoted by $A[\alpha \mapsto B]$, that replaces the free 
  occurrences of variable $\alpha$ in $A$ by $B$, while avoiding the 
  capture of any bound variable in $A$. 
%We consider terms that result after the renaming of bound variables 
%  to be identical to each other. 
When needed, we assume that $\alpha$-equivalence is applied at will 
to avoid the clashing of free variables.

\begin{figure}[t]
  \centering
  \drules[Typ]{$\Gamma \vdash e : \textit{A}$}{Typing}{int, var,
    abs, app, cast}
  
  \drules[Cast]{$\Delta \ottsym{;} \mathbb{E} \vdash \textit{A} \hookrightarrow \textit{B} : c$}{Type Casting}{id, arrow, unfold, fold, seq, var, fix}
  \caption{Typing and type cast rules for \name.}
  \label{fig:typing}
\end{figure}

\subsection{Typing}
\label{subsec:typing}
The top of Figure~\ref{fig:typing} shows the typing rules for \name.
Most rules are standard except for the typing rule for type casting
  (\rref{Typ-cast}). This rule replaces the standard folding and
  unfolding rules for iso-recursive types, as we explained in \S\ref{subsec:results}.
\Rref{Typ-cast} relies on the type casting rules shown at the bottom of Figure~\ref{fig:typing}.
In the type casting judgment $\Delta \ottsym{;} \mathbb{E} \vdash \textit{A} \hookrightarrow \textit{B} : c$,
  $\Delta$ is the type context used to ensure that all the types
  in the cast derivation are well-formed. 
  $\mathbb{E}$ tracks of the
  cast variables $\iota$ that appear in $c$ and the cast operator
  that they are associated with.
New cast variables are introduced when a fixpoint cast is encountered,
  as shown in \rref{Cast-fix}, which gives us the ability to encode
  the coinductive reasoning in equi-recursive equalities.
The cast operator $c$ in the type casting relation essentially
  describes the derivation of a judgment.
Our type casting rules, ignoring the cast variables and operators, 
  are very similar to the type equality rules
  in \citeauthor{brandt1998coinductive}'s axiomatization of type equality.
Despite some subtle differences, which we will discuss in \S\ref{subsec:typingequiv},
  our type casting rules are sound and complete with respect to
  their type equality rules.
\begin{theorem}[Soundness and completeness of type casting]
    \label{thm:cast-sound-complete}
    For any types $A$ and $B$, $\cdot \vdash A \doteq B$ if and only if
    there exists a cast operator $c$ such that $\cdot \ottsym{;} \cdot \vdash A \hookrightarrow B : c$.    
\end{theorem}

\begin{figure}[t]
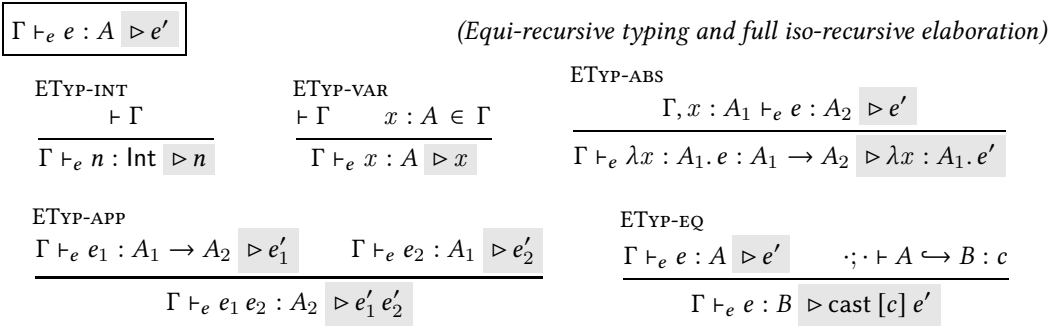

  \centering
  \drules[ETyp]{$\Gamma \vdash_e e : A \grayboxm{\rhd \, e'}$}{Equi-recursive typing and full iso-recursive elaboration}{int, var, abs, app, eq}
  \caption{An equivalent equi-recursive typing system and elaboration rules to \name.}
  \label{fig:equi}
\end{figure}

\paragraph{Equivalence to a calculus with equi-recursive typing.}
The only difference between the equi-recursive typing rules
and \name's typing rules is replacing type casting in \rref{Typ-cast}
with a type equality relation. 
Therefore, we can give an alternative definition of the equi-recursive typing rules
in Figure~\ref{fig:equi}. The gray parts are used to generate a
term in \name, and can be ignored for understanding the equi-recursive
typing rules.
The standard equi-recursive type equality
  relation in \rref{Typ-eq} is replaced by the type casting 
  relation in \rref{ETyp-eq}.
Since the two relations are equivalent by Theorem~\ref{thm:cast-sound-complete},
  the typing rules in Figure~\ref{fig:equi} are equivalent to the 
  standard equi-recursive typing rules. 

\begin{theorem}[Equivalence of alternative equi-recursive typing]
    \label{thm:equi-typing-equiv}
    For any expression $e$ and type $A$,
    $\Gamma \vdash_{e} e : A$ in the standard equi-recursive typing
    with \rref{Typ-eq} if and only if 
    there exists a full iso-recursive term $e'$ such that
    $\Gamma \vdash_{e} e : A \rhd e'$ using the rules in Figure~\ref{fig:equi}.
\end{theorem}
  
Our alternative formulation of equi-recursive typing also provides a way to elaborate equi-recursive terms
  into full iso-recursive terms, as shown by the gray colored parts 
  in Figure~\ref{fig:equi}. 
The elaboration is type-directed, by inserting the appropriate casts
  where a type equality is needed
  following the typing derivation of the equi-recursive terms.
The interesting point here is
  that, by replacing \citeauthor{brandt1998coinductive}'s type
  equality relation with our type casting relation, we obtain a cast
  $c$, which can be viewed as evidence of the type transformation
  from type $A$ into type $B$. Then, we use $c$ in an explicit cast in
  the elaborated \name term,
  which will trigger the type transformation in \name.
Every well-typed equi-recursive term can be elaborated into a full iso-recursive term,
  and every full iso-recursive term can be erased to an equi-recursive term
  that has the same type.
It follows that the full iso-recursive typing rules are sound and
  complete with respect to equi-recursive types:

\begin{theorem}[Equi-recursive to full iso-recursive typing]
  \label{thm:equi-to-full-iso-typing}
  For any expressions $e$, $e'$ and type $A$,
  % $\Gamma \vdash_{e} e : A$ then there exists a full
  % iso-recursive term $e'$ such that
  % elaborating $e$ to a full iso-recursive term $e'$ by
   if $\Gamma \vdash_{e} e : A \rhd e'$ then
   $\Gamma \vdash e' : A$.
\end{theorem}

\begin{theorem}[Full iso-recursive to equi-recursive typing]
  \label{thm:full-iso-to-equi-typing}
  For any expressions $e$ and type $A$,
  if $\Gamma \vdash e : A$ then
  $\Gamma \vdash_{e} |e| : A \rhd e$.
\end{theorem}

% \begin{theorem}[Soundness of elaboration]
%   \label{thm:soundness-elaboration}
%   For any expression $e$ and type $A$, if 
%   $\Gamma \vdash_{e} e : A \rhd e'$ then
%   $\Gamma \vdash_{e} e : A$ in equi-recursive typing and
%   $\Gamma \vdash e' : A$ in the full iso-recursive typing.
% \end{theorem}

\noindent In the theorem above, the full iso-recursive expressions can be erased to the equi-recursive expressions
  by removing the casts.
The erasure operation $|e|$ is defined as follows:
$$
\begin{array}{llcll}
  |\textsf{n}| & = \textsf{n} &\qquad & |x| & = x \\
  |e_1~e_2| & = |e_1|~|e_2| &\qquad & |\lambda x:A .~ e| & = \lambda x:A .~ |e| \\
  |\caste{c}{e}| & = |e|
\end{array}
$$

Moreover, our elaboration achieves the round-tripping property -- 
  elaborating an equi-recursive term into a full iso-recursive term
  and then erasing the casts will get back the original
  equi-recursive term.
% In other words, the casts we introduce are \emph{computationally irrelevant}
%   and can be erased without affecting the behavior of the terms.
This is not the case for previous work in relating recursive types~\cite{abadi1996syntactic}, 
  in which computationally relevant coercions are inserted as term-level
  functions and erasing the unfold/fold annotations do not recover the original term.
The round-tripping property is crucial for a simple proof of the 
  behavioral equivalence between the two systems, which we discuss next.

\begin{theorem}[Round-tripping of the encoding]
  \label{thm:round-trip}
  For any expression $e$, $e'$ and type $A$, 
  if $\Gamma \vdash_{e} e : A \rhd e'$, then
  $|e'| = e$.
\end{theorem}

\subsection{Semantics}\label{subsec:semantics}

Figure~\ref{fig:red} shows the reduction rules for \name.
In addition to the standard reduction rules for the simply typed lambda calculus,
  we add the reduction rules for the cast operators.
Our reduction rules are call-by-value.
The inner expressions of the cast operators are reduced first (\rref{Red-cast}).
Then, based on different cast operators, the cast operator is pushed into the expression
  in various ways.
For identity casts, the cast operator is simply erased (\rref{Red-cast-id}).
Arrow casts are values, but when they are applied to an argument,
  the cast operator is pushed into the argument
  (\rref{Red-cast-arr}).
Note that the cast operator needs to be 
  reversed when pushed into the function argument in order
  to ensure type preservation after the reduction.
The reverse operation is defined by analyzing the structure of $c$ as follows:
$$
\begin{array}{llll}
  \neg\, \iota &= \iota & 
  \neg\, \cid &= \cid \\
  \neg\, \cfold{A} &= \cunfold{A} & 
  \neg\, \cunfold{A} &= \cfold{A} \\
  \neg\, (c_1 \to c_2) &= (\neg\, c_1) \to (\neg\, c_2) & 
  \neg\, (\cseq{c_1}{c_2}) &= \cseq{(\neg\, c_2)}{(\neg\, c_1)} \\
  \neg\, (\cfix{\iota}{c}) &= \cfix{\iota}{\neg\, c}  
\end{array}
$$
A single sequential cast is split into two separate casts (\rref{Red-cast-seq}),
  so that the sub-components can be reduced independently.
Fold casts are values, but can be eliminated by an outer unfold
cast (\rref{Red-cast-elim}). Thus, \rref{Red-cast-elim} corresponds to
the traditional fold/unfold cancellation rule used in calculi with
conventional iso-recursive types.
Finally, fixpoint casts are reduced by unrolling the fixpoint (\rref{Red-cast-fix}).

\begin{figure}[t]
    \centering
    \drules[Red]{$e \hookrightarrow e'$}{Reduction}{beta, appl, appr, cast, castXXid, castXXarr , castXXseq, castelim,  castXXfix}
    \caption{Reduction rules.}
    \label{fig:red}
\end{figure}

The addition of the push rules for the cast operators is necessary for the
  type soundness of \name, since the cast rules are necessary to
  preserve types. \name is type sound, proved with the usual
  preservation and progress theorems:

\begin{theorem}[Progress]
    \label{thm:progress}
    For any expression $e$ and type $A$, if $\cdot \vdash e : A$ then either
    $e$ is a value or there exists an expression $e'$ such that $e \hookrightarrow e'$.
\end{theorem}

\begin{theorem}[Preservation]
    \label{thm:preservation}
    For any expression $e$ and type $A$, if $\cdot \vdash e : A$ and $e \hookrightarrow e'$ then
    $\cdot \vdash e' : A$.
\end{theorem}

\paragraph{Equivalence to the equi-recursive dynamic semantics}
As explained in \S\ref{subsec:results}, the reduction rules
for the cast operators are computationally irrelevant.
Therefore, they can be erased from expressions without
  affecting the behavior of the expressions.
We can obtain the following theorem easily:

\begin{theorem}[Full iso-recursive to equi-recursive behavioral preservation]
    \label{thm:full-iso-to-equi-semantics}
    For any expression $e$, if $e \hookrightarrow^{*} v$ then $|e| \hookrightarrow_{e}^{*} |v|$.
\end{theorem}

\noindent Here $\hookrightarrow_{e}$ is the reduction relation in the equi-recursive setting,
  which is basically defined by a subset of the reduction rules in Figure~\ref{fig:red}
  (\rref{Red-beta,Red-appl,Red-appr}).
We use $\hookrightarrow^{*}$ to denote the reflexive, transitive closure of the 
  reduction relation.
The other direction of the behavioral preservation also holds, but 
  only applies to well-typed expressions and  relies on 
  the elaboration process defined in Figure~\ref{fig:equi}.
The proof of this direction is also more involved, and we will
  detail it in \S\ref{subsec:behavior}.
To summarize, the two systems are behaviorally equivalent, in terms
  of both termination and divergence behavior:  
\begin{theorem}[Behavioral equivalence]
  \label{thm:behavioral-equiv}
  For any expression $e$, $e'$ and type $A$, if $\cdot \vdash_{e} e : A \rhd e'$, then
  \begin{enumerate}
    % erase_reduction
    \item $e \hookrightarrow^{*}_{e} v$ if and only if there exists $v'$ such that $e' \hookrightarrow^{*} v'$ and $|v'| = v$.
    \item $e$ diverges if and only if $e'$ diverges.
    % By divergence we mean that the expression cannot be eventually evaluated to a value.
  \end{enumerate}
\end{theorem}

%%%%%%%%%%%%%%%%%%%%%%%%%%%%%%%%%%%%%%%%%%%%%%%%%%%%%%%%%%%%%%
% Section 4 Metatheory of full iso-recursive types
%%%%%%%%%%%%%%%%%%%%%%%%%%%%%%%%%%%%%%%%%%%%%%%%%%%%%%%%%%%%%%

\section{Metatheory of full iso-recursive types}
\label{sec:metatheory}

In this section we discuss the key proof techniques and results
  in the metatheory of \name.
The metatheory covers three components:
  type soundness of \name (Theorem~\ref{thm:progress} and \ref{thm:preservation}), 
  the typing equivalence between \name and equi-recursive types
    (Theorem~\ref{thm:equi-typing-equiv},~\ref{thm:equi-to-full-iso-typing} and \ref{thm:full-iso-to-equi-typing}),
  and the behavioral equivalence between \name and equi-recursive types
    (Theorem~\ref{thm:behavioral-equiv}).

\subsection{Type Soundness}

\paragraph{Progress}

For \name we need to ensure that the definition of value and the reduction rules,
  in particular the push rules for type casting, are complementary to each other,
  i.e. a cast expression is either a value or can be further reduced.
The definition of value has been discussed in \S\ref{subsec:syntax}.
In \name now there are two canonical forms for a value with function types ($A_1 \to A_2$):
  lambda abstractions $(\lambda x:A .~ e)$ and arrow casts $(\caste{c_1 \to c_2}{v})$.
Therefore in the progress proof for function applications ($e_1~e_2$),
  we need to consider one extra case when $e_1$ is an arrow cast.
We push the cast operator further by \rref{Red-cast-arr} as a reduction step
  in this case to complete the proof.

\paragraph{Preservation}

The preservation proof is standard by first doing induction on the 
  typing derivation $\cdot \vdash e : A$ and then induction 
  on the reduction relation $e \hookrightarrow e'$.
The interesting cases are when the reduction rule is a push rule.
Most cases of the push rules are straightforward,
  by inversion on the type casting relation and then
  reconstructing the casting derivation for the reduced expression.
Two tricky cases require extra care:
  the push rules for arrow cast (\rref{Red-cast-arr}) and the fixpoint cast (\rref{Red-cast-fix}).

In the \rref*{Red-cast-arr} case, 
  by inversion on the typing derivation of 
  $\cdot \vdash (\caste{c_1\to c_2}{v_1})~v_2 : A_2$
  we know that $v_1$ has a function type $B_1 \to B_2$,
  $v_2$ has type $A_1$, and $B_1 \to B_2$ can be cast to $A_1 \to A_2$ by $c_1 \to c_2$,
  which in turn implies that $B_1 \hookrightarrow A_1 : c_1$ and $B_2 \hookrightarrow A_2 : c_2$.
In order to show that the type of the reduced expression
  $(\caste{c_2}{(v_1 ~( \caste{\neg c_1}{v_2}))})$ is still $A_2$,
  we need to prove $A_1 \hookrightarrow B_1 : \neg c_1$.
This goal can be achieved by Lemma~\ref{lem:cast-reverse} below,
  which is proved by induction on the type casting relation.
The analysis of \rref{Red-cast-arr} also
  shows that it is necessary to insert the reverse operation on the cast operator $c_1$
  to ensure the preservation of \name.
\begin{lemma}[Reverse of casting]
    \label{lem:cast-reverse}
    For any types $A$ and $B$, and casting operators $c$, if
    $\cdot \ottsym{;} \cdot \vdash A \hookrightarrow B : c$ then
    $\cdot \ottsym{;} \cdot \vdash B \hookrightarrow A : \neg\, c$.
\end{lemma}

As for the \rref*{Red-cast-fix} case, 
  the reduction rule unfolds the fixpoint cast $(\cfix{\iota}{c})$ to
  $(c [\iota \mapsto (\cfix{\iota}{c}) ])$.
By inversion on the type casting relation 
  $\cdot;\cdot\vdash A \hookrightarrow B : \cfix{\iota}{c}$,
  we know that 
\begin{equation}
  \cdot;\iota: A \hookrightarrow B \vdash A \hookrightarrow B : c
  \label{eqn:fixpoint-cast}
\end{equation}
Essentially the cast operator $(\cfix{\iota}{c})$
  and its unrolling $(c [\iota \mapsto (\cfix{\iota}{c}) ])$
  should represent the same proof.
The type casting judgement (\ref{eqn:fixpoint-cast}) can be interpreted as:
  if we know that there is a cast variable $\iota$ that can cast $A$ to $B$,
  then we can cast $A$ to $B$ by $c$, using the cast variable $\iota$.
Since we already know that $\cfix{\iota}{c}$ can do the same job
  as $\iota$ in casting $A$ to $B$,
  it should be safe to replace $\iota$ with $\cfix{\iota}{c}$ in the cast operator $c$,
  and show that $\cdot;\cdot\vdash A \hookrightarrow B : c [\iota \mapsto (\cfix{\iota}{c}) ]$.
This idea can be formalized by the following cast substitution lemma,
  which is proved by induction on the type casting relation of $c_1$.

\begin{lemma}[Cast substitution]
    \label{lem:cast-subst}
    For any contexts $\Gamma$, $\mathbb{E}$, types $A$, $B$, $C$, $D$,
    cast operators $c_1$, $c_2$ and cast variable $\iota$, if
    $\Gamma \ottsym{;} \mathbb{E} \vdash A \hookrightarrow B : c_1$, and
    $\Gamma \ottsym{;} \mathbb{E}, \iota : A \hookrightarrow B \vdash C \hookrightarrow D : c_2$ then
    $\Gamma \ottsym{;} \mathbb{E} \vdash C \hookrightarrow D : c_2 [ \iota \mapsto c_1 ]$.
\end{lemma}

% The derivation tree seems taking too much space
\begin{comment}
\begin{center}
\def\defaultHypSeparation{\hskip.05in}
\begin{prooftree}
  \AxiomC{$v_1: B_1 \to B_2$}
  \AxiomC{$B_1 \hookrightarrow A_1 : c_1$}
  \AxiomC{$B_2 \hookrightarrow A_2 : c_2$}
  \RightLabel{\rref*{TCast-arrow}}
  \BinaryInfC{$B_1\to B_2 \hookrightarrow A_1\to A_2 : c_1 \to c_2 $}
  \RightLabel{\rref*{Typ-cast}}
  \BinaryInfC{$\caste{c_1 \to c_2}{v_1} : A_1 \to A_2$}
  \AxiomC{$v_2: A_1$}
  \LeftLabel{\rref*{Typ-app}}
  \insertBetweenHyps{\hskip -20pt}
  \BinaryInfC{$(\caste{c_1\to c_2}{v_1})~v_2 : A_2$}
\end{prooftree}

  \begin{prooftree}
    \AxiomC{$v_1: B_1 \to B_2$}
    \AxiomC{$v_2: A_1$}

    % \def\extraVskip{1pt}
    % \AxiomC{Lemma~\ref{lem:cast-reverse}}
    % \noLine
    \AxiomC{$B_1 \hookrightarrow A_1 : c_1$}
    \RightLabel{Lemma~\ref{lem:cast-reverse}}
    \UnaryInfC{$A_1 \hookrightarrow B_1 : \neg c_1 $}
    % \def\extraVskip{3pt}
    
    \RightLabel{\rref*{Typ-cast}}
    \BinaryInfC{$\caste{\neg c_1}{v_2} : B_1 $}
    \RightLabel{\rref*{Typ-app}}
    \insertBetweenHyps{\hskip -8pt}
    \BinaryInfC{$v_1 ~( \caste{\neg c_1}{v_2}) : B_2 $}
    \AxiomC{$B_2 \hookrightarrow A_2 : c_2$}
    \insertBetweenHyps{\hskip -30pt}
    \LeftLabel{\rref*{Typ-cast}}
    \BinaryInfC{$\caste{c_2}{(v_1 ~( \caste{\neg c_1}{v_2}))} : A_2$}
  \end{prooftree}
\end{center}
\end{comment}

\subsection{Typing Equivalence}
\label{subsec:typingequiv}

As discussed in Section~\ref{subsec:typing},
  the key to the typing equivalence between full iso-recursive types
  and equi-recursive types is to show our type casting rules are
  equivalent to \citeauthor{brandt1998coinductive}'s type equality rules 
  (Theorem~\ref{thm:cast-sound-complete}).
This section focuses on the proof of this theorem.

Most of our type casting rules, ignoring the cast variables and operators,
  are very similar to their type equality rules,
  so the proof for these cases is straightforward.
For instance, the treatment of coinductive reasoning by introducing new
  premises for function types in our \rref{Cast-fix} is exactly the same
  treatment as their \rref{Tye-arrfix}.
We discuss the only two differences below.

\paragraph{Arrow cast for type soundness} 
In addition to transforming function types with a fixpoint cast using \rref{Cast-fix}
  as \citeauthor{brandt1998coinductive} did in \rref{Tye-arrfix}, 
  we also allow function types to be cast
  without a fixpoint cast as well, as shown in \rref{Cast-arr}.
This is a harmless extension, since one can always wrap an arrow cast with
  a dummy fixpoint, which does not use the fixpoint variable in the body.
However, having this rule is essential to the type soundness of \name.
By \rref{Cast-fix}, all the fixpoint casts in 
  well-typed expressions are in the form of $\cfix{\iota}{c_1 \to c_2}$.
During the reduction, we need to unroll those fixpoint casts
  using \rref{Red-cast-fix}
  to a bare arrow cast in the form of $c'_1 \to c'_2$,
  which cannot be typed without the \rref{Cast-arr}.
In other words, while casts of arrows are values, casts of fixpoints
  are not values. 
Due to this difference we separate the two rules and prove that the extension
  does not affect the soundness and completeness of our type casting rules.

\paragraph{Removing the symmetry rule from equality} 
The other difference is that
  our rules do not include a symmetry rule for type casting,
  since it is hard to interpret symmetry in the operational semantics.
As a result, changing the type cast context 
  from $\cdot$ to a universally quantified $\mathbb{E}$ in 
  Lemma~\ref{lem:cast-reverse} will not work,
  i.e. \rref{Tye-symm} in its general form, with 
  the same list of assumptions $H$
  in both the premise and conclusion,
  is not admissible in our type casting relation.
The reason is that invalid assumptions may exist in the list $H$,
  which are not derivable by the type casting rules.
For example,
$
\tint \doteq \tint \to \tint \vdash \tint \to \tint \doteq \tint
$
  is a valid judgement using \rref{Tye-symm,Tye-assump},
  but cannot be derived from our type casting rules.

Nevertheless we can still prove that our system is complete 
  to \citeauthor{brandt1998coinductive}'s equality, 
  when the initial environment is empty.
The idea is that starting from an empty assumption list,
  one can always replace the use of \rref{Tye-symm} in the derivation 
  with a complete derivation that redoes the proof goal in the symmetry way 
  to obtain a derivation without using \rref{Tye-symm}.
The replacement is feasible since when the initial environment is empty,
  all the type equalities introduced to the environment are guaranteed
  to be derivable from an empty assumption list
  by the type casting rules.
Interested readers can refer to our Coq formalization for the details
  of the proof.

% \begin{theorem}[Full Iso to Equi]
%     \label{thm:full-iso-to-equi}
% \end{theorem}

% \begin{lemma}[Soundness of TypCast]
%     \label{lem:cast-sound}
% \end{lemma}

% \begin{theorem}[Full Iso to Equi]
%     \label{thm:full-iso-to-equi}
% \end{theorem}

% \begin{lemma}[Completeness of TypCast]
%     \label{lem:cast-complete}
% \end{lemma}

\subsection{Behavioral Equivalence}
\label{subsec:behavior}

To prove Theorem~\ref{thm:behavioral-equiv} it suffices to show one
  of the two propositions in the theorem. 
Since the type soundness of \name ensures
  that a well-typed term does not get stuck -- it can either diverge
  or reduce to a value, we only need to show the preservation
  of termination behavior and the preservation of divergence behavior
  can then be proved by contradiction.
The ``if'' direction of the theorem (from full-iso recursive reduction
  to equi-recursive reduction) is easy, directly from the behavioral
  preservation property of the erasure function (Theorem~\ref{thm:full-iso-to-equi-semantics}).
The ``only if'' direction (from equi-recursive reduction to full-iso recursive reduction)
  is more involved, and we prove it by induction on the length of
  reduction steps.
It suffices to show the following lemma, which states that one step
  of equi-recursive reduction can be simulated by several steps
  of full iso-recursive reduction.

% reduction_e2i
\begin{lemma}[Simulation of equi-recursive reduction]
    \label{lemma:equi-to-full-iso-step}
    For any expressions $e_1$, $e'_1$, $e_2$ and type $A$, if $\cdot \vdash_{e} e_1 : A \rhd e_1'$ 
    and $e_1 \hookrightarrow_{e} e_2$, then
    there exists $e_2'$ such that $e_1' \hookrightarrow^{*} e_2'$ 
    and $\cdot \vdash_{e} e_2 : A \rhd e_2'$.
\end{lemma}

The proof of Lemma~\ref{lemma:equi-to-full-iso-step} is done by first induction on the
  equi-recursive reduction relation $e_1 \hookrightarrow_{e} e_2$, and
  then induction on the elaboration relation $\cdot \vdash_{e} e_1 : A \rhd e_1'$.
Most of the cases are straightforward, by applying the induction hypothesis 
  and using the congruence lemmas
  below to construct the reduction steps.
\begin{lemma}[Congruence lemma of full iso-recursive reduction] \,
    \label{lem:cong-red}
    \begin{enumerate}
      \item \label{lem:cong-red1} If $e_1 \hookrightarrow^* e_1'$, then $e_1~e_2 \hookrightarrow^* e_1'~e_2$.
      \item \label{lem:cong-red2} If $e_2 \hookrightarrow^* e_2'$, then $v_1~e_2 \hookrightarrow^* v_1~e_2'$.
      \item If $e \hookrightarrow^* e'$, then $\caste{c}{e} \hookrightarrow^* \caste{c}{e'}$.
    \end{enumerate}
\end{lemma}
The tricky case is when $e_1 \hookrightarrow_{e} e_2$ is a beta reduction (case \rref*{Red-beta}).
By inversion on the reduction relation, 
  we know that $e_1$ is a function application ($e_1 = (\lambda x:A_1.e_0)~v_1$)
  and $e_2$ is ($e_0[x\mapsto v_1]$).
By induction on the elaboration relation $\cdot \vdash_{e} e_1 : A \rhd e_1'$,
  case \rref*{ETyp-eq} can be proved using the induction hypothesis.
We consider case \rref*{ETyp-app}, where
\begin{equation}
  \label{eqn:tricky-case}
  \cdot\vdash \lambda x:A_1.e_0 : A_1 \to A \rhd e_3' \quad \text{ and } \quad
  \cdot\vdash v_1 : A_1 \rhd e_4'  
\end{equation}
However, we do not know the exact form of $e_3'$ and $e_4'$,
  since many different casts can be inserted in the elaboration derivation
  using \rref{ETyp-eq},
  and the current induction hypothesis cannot deal with this.
To address this issue, we first prove a lemma showing that
  any full iso-recursive terms elaborated from an equi-recursive \emph{value}
  can always be further reduced to a value in the full iso-recursive setting.
\begin{lemma}[Reductions of full iso-recursive terms from equi-recursive values]
    \label{lem:equi-to-full-iso-value}
    For any expression $e$ and type $A$, if $\cdot \vdash_{e} v : A \rhd e'$, then
    there exists $v'$ such that $e' \hookrightarrow^{*} v'$ and $\cdot \vdash_{e} v : A \rhd v'$.
\end{lemma}
Lemma~\ref{lem:equi-to-full-iso-value} can be proved by induction on the 
  typing derivation $\cdot \vdash_{e} v : A \rhd e'$ and using
  the congruence lemma (Lemma~\ref{lem:cong-red}).
By applying Lemma~\ref{lem:equi-to-full-iso-value} to (\ref{eqn:tricky-case})
  in the tricky case of Lemma~\ref{lemma:equi-to-full-iso-step}, 
  we can first show that both $e_3'$ and $e_4'$ can be further 
  reduced to a value.
Moreover, the value of evaluating $e_3'$ preserves
  the function type, so it must be one of the two canonical forms:
  a lambda abstraction $(\lambda x:A_1. e'_0 )$ or an arrow cast $(\caste{c_1\to c_2}{v'_1} )$, 
  and then we can construct 
  the reduction steps for $(e'_3~e'_4)$  as shown below:
$$
\begin{array}{ll@{}l@{}ll}
                  &  e'_3~e'_4  \\
\hookrightarrow^* & (\lambda x:A_1. e'_0 ) ~e'_4 &
\text{\quad or\quad} & (\caste{c_1\to c_2}{v'_1} ) ~e'_4  &
  \text{(Lemma~\ref{lem:equi-to-full-iso-value} and \ref{lem:cong-red}(\ref{lem:cong-red1}))} \\
\hookrightarrow^* & (\lambda x:A_1. e'_0 ) ~v'_2 & 
\text{\quad or\quad} & (\caste{c_1\to c_2}{v'_1} ) ~v'_2 &
  \text{(Lemma~\ref{lem:equi-to-full-iso-value} and \ref{lem:cong-red}(\ref{lem:cong-red2}))} \\
\hookrightarrow & e'_0 [x \mapsto v'_2] & 
\text{\quad or\quad} & (\caste{c_2}{(v'_1~(\caste{\neg c_1}{v'_2}))}) &
  \text{(\Rref{Red-beta} or \rref*{Red-cast-arr}) } \\
\end{array}
$$

Now we are left to prove the second goal of 
  Lemma~\ref{lemma:equi-to-full-iso-step},
  that is, the result of the reduction constructed
  above can be derived from the elaboration relation, i.e.
$$
\cdot \vdash_e e_0[x\mapsto v_1] : A \rhd e'_0 [x \mapsto v'_2]
\quad \text{or} \quad
\cdot \vdash_e e_0[x\mapsto v_1] : A \rhd (\caste{c_2}{(v'_1~(\caste{\neg c_1}{v'_2}))})
$$
  % ($\cdot \vdash_{e} e_2 : A \rhd e_2'$).
The latter case for \rref{Red-cast-arr} follows from the
  induction hypothesis.
The first case for \rref{Red-beta} can be proved by
  the following substitution lemma for the elaboration relation.

\begin{lemma}[Substitution lemma for elaboration]
    \label{lem:subst-elab}
  For any typing context $\Gamma$, expressions $e_1$, $e_1'$, $e_2$, $e_2'$ and types $A$, $B$, if
  $\Gamma,  x : A  \vdash_{e} e_1 : B \rhd e_1'$ and 
  $\Gamma \vdash_{e} e_2 : A \rhd e_2'$,
  then $\Gamma  \vdash_{e} e_1[x \mapsto e_2] : B \rhd e_1'[x \mapsto e_2']$.
\end{lemma}

With Lemma~\ref{lem:cong-red},~\ref{lem:equi-to-full-iso-value} and \ref{lem:subst-elab},
  we complete the proof of Lemma~\ref{lemma:equi-to-full-iso-step},
  and the behavioral preservation theorem (Theorem~\ref{thm:behavioral-equiv}) follows.
Compared to the behavioral equivalence proof in \citeauthor{abadi1996syntactic}'s work,
  we show that it is much more straightforward to prove the behavioral equivalence between
  full iso-recursive types and equi-recursive types, and our proof for \name 
  is completely mechanized in Coq, without relying on any conjectures or axioms.

%%%%%%%%%%%%%%%%%%%%%%%%%%%%%%%%%%%%%%%%%%%%%%%%%%%%%%%%%%%%%%
% Section 5 Towards Subtyping
%%%%%%%%%%%%%%%%%%%%%%%%%%%%%%%%%%%%%%%%%%%%%%%%%%%%%%%%%%%%%%

\section{Recursive Subtyping}

% Section 5 Towards Subtyping

In this section we show that our results in the previous sections can be
extended to a calculus with subtyping called \nameS.

\subsection{A Calculus with Subtyping}

Adapting the results in Section~\ref{sec:calculus} to a calculus with subtyping
requires only a few changes. In terms of types, we add a top
type ($\ttop$). Expressions and values remain the same.

\begin{figure}
  \drules[Sub]{$\Sigma \vdash A \le_{\oplus} B$}{Equi-recursive/Iso-recursive Subtyping}
  {top, int, eq, var, self, trans, arrow, rec}
  \caption{\citeauthor{amadio1993subtyping}'s
    equi-recursive and iso-recursive subtyping rules. 
    }
  \label{fig:subtyping}
\end{figure}

\paragraph{Subtyping}
The equi-recursive and iso-recursive subtyping rules 
 that we use in this section are based on the Amber rules~\cite{amadio1993subtyping}, as shown in Figure~\ref{fig:subtyping}.
The subtyping rules use a special environment $\Sigma$, which
  tracks a set of pairs of type variables that are 
  assumed in the subtyping relation, as explained in \S\ref{subsec:subtyping}.
We use $\le_{\oplus}$ parameterized by a metavariable $\oplus \in \{i, e\}$ 
  to denote the subtyping rules for both relations: $i$ denotes
  iso-recursive subtyping, and $e$ denotes equi-recursive subtyping.
We use $\le_{e}$ to denote the subtyping rules (\rref{Sub-eq,Sub-trans}) 
  that only apply to equi-recursive types,
  and $\le_{i}$ to denote the subtyping rules (\rref{Sub-int,Sub-self})
  that only apply to iso-recursive types.
\Rref{Sub-eq} embeds the equi-recursive equality relation in Figure~\ref{fig:ac-tyeq}
  into the subtyping relation, so is only present in
  equi-recursive subtyping.
For the iso-recursive subtyping relation, we choose the variant of the Amber rules
  presented by \citet{zhou2022revisiting}, which replaces the built-in
  reflexivity with the more primitive \rref{Sub-int,Sub-self} and removes transitivity \rref{Sub-trans}
  from the original Amber rules.
\citeauthor{zhou2022revisiting} discussed the technical challenges of having 
  reflexivity and transitivity built-in in the iso-recursive subtyping relation, 
  and showed that they are admissible from the other rules.
\begin{lemma}[Reflexivity of iso-recursive subtyping]
    \label{lemma:reflexivity-iso}
    If $A$ is a closed type, then $\Sigma \vdash A \le_{i} A$.
\end{lemma}
\begin{lemma}[Transitivity of iso-recursive subtyping]
  \label{lemma:transitivity-iso}
  If $\cdot \vdash A \le_{i} B$ and $\cdot \vdash B \le_{i} C$, then $\cdot \vdash A \le_{i} C$.
\end{lemma}

In Lemma~\ref{lemma:reflexivity-iso}, the assumption that $A$ is a closed type 
  can be implied by the fact that the free variable sets of
   $A$ and $B$ in the subtyping relation
  $\Sigma \vdash A \le_{i} B$ are disjoint, which is also a side condition in
  \citeauthor{amadio1993subtyping}'s equi-recursive Amber rules.
As for the transitivity, Lemma~\ref{lemma:transitivity-iso} only holds
  when the environment is empty.
Otherwise, one may also get into problematic subtyping relations, 
  as discussed by \citeauthor{zhou2022revisiting}.
For the reasons above, we choose to use the variant of iso-recursive Amber
  rules without built-in reflexivity and transitivity in this section.

\paragraph{Typing and Reduction}
As for the typing rules, we extend the full iso-recursive type system in 
  Figure~\ref{fig:typing} with \rref{Typ-sub}, and the equi-recursive type system 
  in Figure~\ref{fig:equi} with \rref{ETyp-isub,ETyp-bare-sub}, respectively.
\Rref{ETyp-isub} allows the elaboration rule to consider iso-recursive subtyping
  subsumptions, as \rref{Typ-sub} does in the typing rules for \nameS.
\Rref{ETyp-bare-sub} is required for the calculus with equi-recursive subtyping.
Note that \rref{ETyp-bare-sub} only exists in the equi-recursive type system.
The elaboration rule for this case will be discussed later in this section.
There are no changes to the reduction rules.
\begin{displaymath}
  \text{\drule{Typ-sub}}\quad
  \text{\drule{ETyp-isub}}\quad
  \text{\drule{ETyp-bare-sub}  }
\end{displaymath}

\subsection{Type Soundness}
There are no significant technical challenges in extending the type soundness 
  proof to \nameS.
The only part that requires extra care is the preservation lemma, 
  in which we need to show that \rref{Red-castelim} preserves the
  typing in the presence of subtyping.
Let us consider an expression 
  $\caste{\cunfold{\mat{A}}}{(\caste{\cfold{\mat{B}}}{v})}$.
The derivation below shows the typing of this expression.
\begin{center}
\begin{prooftree}
  \AxiomC{$\cdot \vdash v: B[\alpha \mapsto \mat{B}]$}
  \AxiomC{$\ldots$}
  \LeftLabel{\rref*{Typ-cast}}
  \BinaryInfC{$\cdot \vdash \caste{\cfold{\mat{B}}}{v} : \mat{B}$}
  \AxiomC{$\cdot \vdash \mat{B} \le_{i} \mat{A}$}
  \LeftLabel{\rref*{Typ-sub}}
  \BinaryInfC{$\cdot \vdash \caste{\cfold{\mat{B}}}{v} : \mat{A}$}
  % \AxiomC{}
  % \RightLabel{\rref*{Cast-unfold}}
  % \UnaryInfC{$\cdot;\cdot \vdash  \mat{A} \hookrightarrow A[\alpha \mapsto \mat{A}] : \cunfold{\mat{A}} $}
  \AxiomC{$\ldots$}
  \LeftLabel{\rref*{Typ-cast}}
  \BinaryInfC{$\cdot \vdash 
    \caste{\cunfold{\mat{A}}}{(\caste{\cfold{\mat{B}}}{v})} : A[\alpha \mapsto \mat{A}]$}
\end{prooftree}
\end{center}
By inversion we know that after reduction using 
  \rref{Red-castelim}, the result $v$ has the type
  $B[\alpha \mapsto \mat{B}]$, and that $\mat{B} \le_{i} \mat{A}$.
% To prove preservation we need to show $B[\alpha \mapsto \mat{B}] \le_{i}
%   A[\alpha \mapsto \mat{A}]$.
The preservation proof goal for this case can be expressed as the 
  following lemma:
\begin{lemma}[Unfolding lemma]
  \label{lemma:unfolding}
  If $\cdot \vdash \mat{B} \le_{i} \mat{A}$, then $\cdot \vdash B[\alpha \mapsto \mat{B}] \le_{i} A[\alpha \mapsto \mat{A}]$.
\end{lemma}
The proof of this lemma has been shown by
\citet[Corollary 59]{zhou2022revisiting}. They
proposed an alternative formulation of the iso-recursive subtyping
rules, which is equivalent to the iso-recursive Amber rules. 
Therefore, by adopting their results, we complete the type soundness
  proof for \nameS.

\begin{theorem}[Type soundness of \nameS] For any term $e$ and type $A$ in \nameS,
  \label{thm:soundness_sub}
  \begin{enumerate}
    \item (Progress) if $\cdot \vdash e : A$ then either
    $e$ is a value or there exists a term $e'$ such that $e \hookrightarrow e'$.
    \item (Preservation) if $\cdot \vdash e : A$ and $e \hookrightarrow e'$ then
    $\cdot \vdash e' : A$.
  \end{enumerate}
\end{theorem}

% \begin{theorem}[Preservation of \nameS]
%   \label{thm:preservation_sub}
%   For any expression $e$ and type $A$ in \nameS, if $\cdot \vdash e : A$ and $e \hookrightarrow e'$ then
%   $\cdot \vdash e' : A$.
% \end{theorem}

\subsection{Typing equivalence}
Similarly to \name with equality, we can prove that \nameS with iso-recursive
  subtyping is sound and complete with respect to a calculus with
  equi-recursive subtyping.
 The key idea is to encode the equi-recursive subtyping relation
  using a combination of equi-recursive equality and the iso-recursive 
  subtyping relation, as explained in \S\ref{subsec:results}.
The encoding can be justified by the following theorem:
\begin{restatable}[Equi-recursive subtyping decomposition]{theorem}{decompthm}
  \label{thm:equi-to-iso}
  $\cdot \vdash A \le_{e} B$ if and only if there exist types $C_1$ and $C_2$ such that $A \doteq C_1$, $\cdot \vdash C_1 \le_{i} C_2$, and $C_2 \doteq B$.  
\end{restatable}
The soundness direction (i.e. the ``if'' direction) of this lemma is straightforward
  by \rref{Sub-eq,Sub-trans} and the fact that
  $\le_{i}$ is a sub-relation of $\le_{e}$.  
The completeness direction can be derived from 
  \citeauthor{amadio1993subtyping}'s proof of completeness with
  respect to the tree model for the equi-recursive Amber rules.
They proved that for any types $A$, $B$ that are in the tree model 
  interpretation of the equi-recursive subtyping relation, one can
  find types $C_1$ and $C_2$ such that $A \doteq C_1$, $C_1 \le_{e} C_2$, and $C_2 \doteq B$ hold~\cite[Lemma~5.4.1, Lemma~5.4.3]{amadio1993subtyping}.
Moreover, the derivation of $C_1 \le_{e} C_2$ satisfies 
  the \emph{one-expansion property}, which means that in the derivation
  each recursive type is unfolded at most once, informally speaking.
Although this result is expressed as an equi-recursive subtyping relation
  in their conclusion, we can rewrite all the occurrences of
  $C_1 \le_{e} C_2$ with one-expansion property in their proof to 
  an iso-recursive subtyping relation $C_1 \le_{i} C_2$. 
  Every application of \rref{Sub-eq,Sub-trans} in their
  proofs can either be replaced by
  \rref{Sub-rec}, in which the recursive type body does not involve the 
  type variable and unfolds to itself, or 
  by \rref{Sub-self} for two recursive types that are
  syntactically equal up to $\alpha$-renaming.
In other words, \citeauthor{amadio1993subtyping}'s proof of their Lemma~5.4.3
  can be seen as a proof that the iso-recursive subtyping relation is complete
  with respect to the equi-recursive subtyping relation with the one-expansion property,
  because they never use the power of the \rref{Sub-eq,Sub-trans}.

\begin{figure}[t]

\def\defaultHypSeparation{\hskip.05in}
\def\extraVskip{3pt}

  \begin{prooftree}
    
    \AxiomC{$\vdash \tint \le_{e} \tint $}

    \def\extraVskip{0pt}
    \AxiomC{\rref*{Tyeq-unfold}}
    \noLine
    \UnaryInfC{\rref*{Tyeq-contract}}
    \noLine
    \UnaryInfC{$ A_1 \doteq \ttop \to A_2 $}
    \def\extraVskip{3pt}

    \AxiomC{$\vdash \ttop \le_{e} \ttop$}
    \AxiomC{$\mathcal{P}_1$}
    \RightLabel{\rref*{Sub-rec}}
    \UnaryInfC{$\vdash A_2 \le_{e} B$}
    \RightLabel{\rref*{Sub-arrow}}
    \BinaryInfC{$  \vdash \ttop \to A_2 \le_{e} \ttop \to B $}

    \RightLabel{\rref*{Sub-eq,Sub-trans}}
    \BinaryInfC{$\vdash A_1 \le_{e} \ttop \to B$}
    \RightLabel{\rref*{Sub-arrow}}
    \insertBetweenHyps{\hskip -30pt}
    \BinaryInfC{$\vdash A \le_{e} \tint \to \ttop \to B$}
    \def\extraVskip{0pt}
    \AxiomC{\rref*{Tyeq-unfold}}
    \noLine
    \UnaryInfC{$\vdash \tint \to \ttop \to B \doteq B$}
    \def\extraVskip{3pt}

    \insertBetweenHyps{\hskip -60pt}
    \RightLabel{$\begin{array}{@{}l@{}}
      \text{\rref*{Sub-trans}} \\[-2pt]
      \text{and \rref*{Sub-eq}}
    \end{array}$}
    \BinaryInfC{$\vdash A \le_{e} B$}
  \end{prooftree}

  \begin{prooftree}
    \AxiomC{$\vdash \tint \le_{i} \tint $}

    \AxiomC{Lemma~\ref{lemma:reflexivity-iso}}
    \def\extraVskip{1pt}
    \noLine
    \UnaryInfC{$ \vdash \ttop \to A_2 \le_{i} \ttop \to A_2 $}
    \def\extraVskip{3pt}

    \AxiomC{$\vdash \ttop \le_{i} \ttop$}
    % \AxiomC{$\alpha \le \beta \vdash \ttop \to \ttop \to \alpha \le_{e} \tint \to \top \to \beta$}
    
    \AxiomC{$\mathcal{P}_2$}
    \RightLabel{\rref*{Sub-rec}}
    \UnaryInfC{$\vdash A_2 \le_{i} B$}
    \RightLabel{\rref*{Sub-arrow}}
    \BinaryInfC{$  \vdash \ttop \to A_2 \le_{i} \ttop \to B $}
    \RightLabel{Lemma~\ref{lemma:transitivity-iso}}

    % \LeftLabel{\rref*{Sub-eq,Sub-trans}}
    \BinaryInfC{$\vdash \ttop \to A_2 \le_{i} \ttop \to B $}
    \RightLabel{\rref*{Sub-arrow}}

    \insertBetweenHyps{\hskip -20pt}
    \BinaryInfC{$\vdash C \le_{i} D$}
    % \AxiomC{}
    % \RightLabel{\rref*{Cast-unfold}}
    % \UnaryInfC{$\cdot;\cdot \vdash  \mat{A} \hookrightarrow A[\alpha \mapsto \mat{A}] : \cunfold{\mat{A}} $}
    % \RightLabel{\rref*{Sub-eq,Sub-trans}}
    \def\extraVskip{1pt}
    \AxiomC{Lemma~\ref{lemma:reflexivity-iso}}
    \noLine
    \UnaryInfC{$\vdash D \le_{i} D$}
    \def\extraVskip{3pt}

    \RightLabel{Lemma~\ref{lemma:transitivity-iso}}
    \BinaryInfC{$\vdash C \le_{i} D$}
  \end{prooftree}

  $$
  \begin{array}{l}
  \begin{array}{ll}
    A = \tint \to (\mat{\ttop \to \alpha}) & B = \mat{\tint \to \top \to \alpha} \\
    C = \tint \to \ttop \to (\mat{\ttop \to \ttop \to \alpha}) & D = \tint \to \ttop \to (\mat{\tint\to \ttop \to \alpha}) \\
    A_1 = \mat{\ttop \to \alpha}  & A_2 = \mat{\ttop \to \ttop \to \alpha}
  \end{array} \\
  \,\,\,\mathcal{P}_1 \text{ is the judgement } \alpha \le \beta \vdash \ttop \to \ttop \to \alpha \le_{e} \tint \to \top \to \beta \\
  \,\,\,\mathcal{P}_2 \text{ is the judgement } \alpha \le \beta \vdash \ttop \to \ttop \to \alpha \le_{i} \tint \to \top \to \beta
  \end{array}
  $$

  \caption{Illustration of decomposing an equi-recursive subtyping derivation.}
  \label{fig:decomp}
\end{figure}

The idea of this decomposition can be illustrated by an example in Figure~\ref{fig:decomp}.
The upper part of the figure shows a derivation by following
  \citeauthor{amadio1993subtyping}'s subtyping algorithm for equi-recursive subtyping.
Note that there are two applications of \rref{Sub-eq} in the derivation,
  one for expanding the type $\mat{\ttop \to \alpha}$ to $\ttop \to \mat{\ttop \to \ttop \to \alpha}$
  and the other for expanding the type $B$ to its unfolding $\tint \to \top \to B$.
Although \rref{Sub-eq} is applied in the middle of the derivation, 
  we can always lift these uses of \rref{Sub-eq} to the top of the derivation,
  by replacing two types in the conclusion with their more expanded forms.
The lower part of the figure shows such a derivation, in which
  we use $C$ and $D$ to denote (a simplified form of) the expanded types
  obtained from \citeauthor{amadio1993subtyping}'s proof.
A key observation here is that the original structure of the derivation
  is preserved in the new derivation.
To highlight this, we use a dummy application of the reflexivity and
  transitivity lemma to show the
  correspondence between the two derivations.

With the decomposition theorem, we can use the following rule to encode 
  the equi-recursive subtyping relation:
\begin{center}
  \drule{ETyp-sub}  
\end{center}
If one ignores the gray parts in the rule, \rref{ETyp-sub} is
  equivalent to \rref{ETyp-bare-sub}. 
We can first apply Theorem~\ref{thm:cast-sound-complete} to
  rewrite our type casting rules to equi-recursive equalities and
  then use Theorem~\ref{thm:equi-to-iso} to show the equivalence to 
  the equi-recursive subtyping relation.
On the other hand, \rref{ETyp-sub} can be derived from 
  the primitive \rref{ETyp-eq,ETyp-isub} in \nameS.
Therefore, we can conclude that \nameS with iso-recursive subtyping is
  sound and complete with respect to a calculus with equi-recursive subtyping
  in terms of typing.
% In other words, Theorem~\ref{thm:full-iso-to-equi-typing} and Theorem~\ref{thm:equi-to-full-iso-typing}
%   also hold for \nameS with subtyping.

\begin{theorem}[Typing equivalence for \nameS] For any expressions $e$, $e'$ and type $A$,
  \label{thm:typing-equiv-sub}
  \begin{enumerate}
    \item (Soundness) if $\Gamma \vdash e : A$ then  $\Gamma \vdash_{e} |e| : A \rhd e$.
    \item (Completeness) if $\Gamma \vdash_{e} e : A \rhd e'$ then $\Gamma \vdash e' : A$.
    \item (Round-tripping) if $\Gamma \vdash_{e} e : A \rhd e'$, then $|e'| = e$.
  \end{enumerate}
\end{theorem}

\subsection{Behavioral equivalence}
We also show that \nameS with iso-recursive
  subtyping is sound and complete with respect to a calculus with equi-recursive
  subtyping in terms of dynamic semantics.
Since there are no changes to the reduction rules, the proof of 
  \nameS to equi-recursive behavioral equivalence by erasure of cast operators
  remains the same as Theorem~\ref{thm:full-iso-to-equi-semantics}.
The proof of the other direction comes almost for free as well.
We simply follow the same steps as described in \S\ref{subsec:behavior} and
  use the same lemmas and theorems to show the completeness of \nameS in
  preserving the equi-recursive reductions,
  except that during the proof of
  Lemma~\ref{lemma:equi-to-full-iso-step},
  we may need to insert an application of \rref{ETyp-isub} at certain points
  to prove that the encoding is well-typed.
In terms of dynamic semantics, \nameS is equivalent to a calculus with 
  equi-recursive subtyping.

\begin{theorem}[Behavioral equivalence of \nameS]
  \label{thm:behavioral-equiv-sub}
  For any expression $e$, $e'$ and type $A$ in \nameS, if $\cdot \vdash_{e} e : A \rhd e'$, then
  \begin{enumerate}
    \item $e \hookrightarrow^{*}_{e} v$ if and only if there exists $v'$ such that $e' \hookrightarrow^{*} v'$ and $|v'| = v$.
    \item $e$ diverges if and only if $e'$ diverges.
    % By divergence we mean that the expression cannot be eventually evaluated to a value.
  \end{enumerate}
\end{theorem}

\section{Related Work}

Throughout the paper, we have discussed some of the closest related work in detail. 
This section covers additional related work.

\paragraph{Relating iso-recursive and equi-recursive types}
Recursive types were first introduced by \citeauthor{morris1968lambda}, 
  who presented equi-recursive types to model recursive definitions. 
Later on, iso-recursive types were
  introduced~\cite{harper1993type,gunter1992semantics,crary1999recursive}. 
The terms for these two types of recursive formulations were 
  coined by \citeauthor{crary1999recursive}.

Both equi-recursive and iso-recursive types 
  have been applied in various programming language areas. 
Equi-recursive types are used 
  in several contexts, including: session types~\cite{castagna2009foundations,chen2014preciseness,gay2005subtyping,gay2010linear}, 
  gradual typing~\cite{siek2016recursive}, 
  and the foundation of Scala through 
    Dependent object types (DOT)~\cite{amin2016essence,rompf2016type}, among others. 
Iso-recursive types have also been utilized in different calculi and language designs 
  due to their ease of use in type checking~\cite{abadi2012theory,bengtson2011refinement,chugh2015isolate,duggan2002type,lee2015theory,swamy2011secure}.

\citet{urzyczyn1995positive} studied the relationship between positive
  iso- and equi-recursive types, showing their equivalence in typing
  power.  Closer to our work, \citet{abadi1996syntactic} explored
  translating equi-recursive terms to iso-recursive terms using
  explicit coercion functions but did not address the operational
  semantics.  Moreover, their behavioral equivalence argument relies on
  a program logic which was conjectured to be sound. As we have argued
  in Section~\ref{subsec:relating}, the use of explicit coercions has
  important drawbacks. Firstly, it adds significant computational
  overhead to the encoding, making the encoding impractical. Secondly,
  it introduces major complications to reasoning, and also prevents a
  round-tripping property. By using casts, we avoid both of these
  issues, leading to an easier, and fully formalized, way to
  establish behavioral equivalence.
  %The round-tripping property is of practical
  %relevance, because it enables the possibility of source languages
  %that employ some form of equi-recursive typing

Recently, \citet{patrignani2021semantic} examined the contextual equivalence
  between iso- and equi-recursive types, 
  providing a mechanized proof in Coq for fully abstract compilers.
  Their focus was on the compilation 
  from iso-recursive to equi-recursive types.
They proved that the translation from iso-recursive to equi-recursive types,
  by erasing unfold/fold operations,
  is fully abstract with respect to contextual equivalence.
The work also covered the compiler from term-level fixpoints to equi-recursive types,
   but did not explore the translation from equi-recursive to
  iso-recursive types.
In our work,
  we establish the bidirectional equivalence between full iso-recursive and 
  equi-recursive types, taking into account both typing and
  operational semantics.
  Furthermore, in addition to type equality, we also study calculi with
  subtyping, which have not been covered in previous work studying the
  relationship between iso- and equi-recursive typing.
%The reverse direction, from equi-recursive to iso-recursive types, 
% has not been studied in depth.

\paragraph{Subtyping recursive types}
\citet{amadio1993subtyping} were the first to 
  present a comprehensive formal study of subtyping with equi-recursive types.
This work inspired further research that refined and 
  simplified the original study~\cite{brandt1998coinductive,gapeyev2002recursive,danielsson2010subtyping,komendantsky2011subtyping}. 
In particular, \citet{brandt1998coinductive} introduced a fixpoint rule for 
  a coinductive relation within an inductive framework. 
Their rules give rise to a natural operational interpretation 
  of proofs as coercions, as they indicated as future work in their paper.
Our work is inspired by their work, and we formally present an
  operational interpretation of equi-recursive equalities in our paper.
However, instead of using coercions to model the subtyping relation as they suggested,
  we use cast operators to model the equalities between equi-recursive types.
Furthermore, we show that our computationally irrelevant cast operators
  simplify the metatheory and extend to subtyping as well.

Iso-recursive subtyping, notably through the Amber rules introduced 
  by Cardelli~\cite{cardelli1985amber}, has long been used in practice.
The iso-recursive Amber rules, while easy to implement, are difficult
  to reason with formally. The only known direct proof for transitivity of
  subtyping for an algorithmic version of the Amber rules was given by
  \citet{bengtson2011refinement}. However this proof relies on a complex inductive argument
  and was found difficult to formalize in theorem provers
  \cite{backes14union, zhou2022revisiting}.
\citeauthor{zhou2022revisiting}, proposed alternative formulations of
  iso-recursive subtyping, which are equivalent to the Amber rules and
  are also easier to reason with. Their work comes with a
  comprehensive formalization of the metatheory of iso-recursive subtyping.
Our work is based on some of their findings. In particular we reuse their mechanized proof
  of the unfolding lemma to show the type soundness of iso-recursive subtyping,
  but instead apply it in a setting with full iso-recursive
  types. Thus, we extend their work to a more general setting, 
  in terms of typing and operational semantics.

To address the complexities of iso-recursive subtyping, 
  several alternative formulations of iso-recursive subtyping
  have been proposed. 
\citet{hofmann1995positive} introduced a subtyping relation that limits 
  recursive subtyping to covariant types only, making the rules more restrictive 
  than the Amber rules. 
\citet{ligatti2017subtyping} offered a broader subtyping 
  relation for iso-recursive types, allowing a recursive type 
  and its unfolded version to be considered subtypes of each other. 
This approach extends the iso-recursive Amber rules but is still not complete
  with respect to the equi-recursive subtyping, since it does not
  consider types not directly related by unfolding or folding as subtypes. 
Additionally, \citet{rossberg2023mutually} developed a calculus for
  higher-order iso-recursive subtyping, to handle mutually recursive types more effectively.

\paragraph{Mechanizing recursive types}

\citet{danielsson2010subtyping,jones2016mechanical} formalized
  equi-recursive subtyping relations in Agda using 
  a mixed coinduction and induction technique.
\citeauthor{jones2016mechanical} presented a semantic 
  interpretation of subtyping and proved
  that their semantic interpretation is sound with respect 
  to an inductive interpretation of types,
  but they did not lift their results to cover function
  types. Instead, they focused on other constructs like product and sum types.
\citeauthor{danielsson2010subtyping} are closer to our work
  since they also did not consider semantic interpretations,
  but formalized in Agda an alternative equi-recursive
   subtyping relation that allows an explicit transitivity rule
   to be included.
They formally proved that this relation is equivalent to the tree model of subtyping 
  as well as \citeauthor{brandt1998coinductive}'s subtyping relation.
In a similar vein, \citet{komendantsky2011subtyping} showed how to 
  implement mixed coinduction and induction within Coq, formalizing rules 
  that closely resemble those introduced by \citet{danielsson2010subtyping}. 
They also validated their approach against \citeauthor{amadio1993subtyping}'s tree model of subtyping.
\citet{zhou2022revisiting} focused on formalizing Amber-style iso-recursive subtyping 
  in Coq, adding to the understanding of iso-recursive subtyping.
\citet{patrignani2021semantic}, which we have discussed earlier, formalized three calculi in Coq: 
  a simply typed lambda calculus extended with iso-recursive types, 
  equi-recursive types, and term-level fixpoints.
Their work is focused on the translation from iso-recursive to equi-recursive types,
  by erasing unfold/fold operations, and the translation from a
   calculus with term-level fixpoints to a calculus with equi-recursive types.
All of our results are mechanized in Coq, with the exception of 
  the decomposition lemma (Theorem~\ref{thm:equi-to-iso}).
  This lemma is implied from \citeauthor{amadio1993subtyping}'s work, but
   relies on a significant amount of technical machinery, which we
   have not formalized in Coq. Thus we assume it as an axiom in our
   Coq formalization.

\paragraph{Casts for type-level computation}
In this paper, we employ explicit cast operators to represent 
  the transformations between types related by equi-recursive equalities. 
Several studies~\cite{stump2009verified,sjoberg2012irrelevance,kimmell2012equational,
    sjoberg2015programming,cretin2014erasable,sulzmann2007system,
    gundry2013type,weirich2017specification,yang2019pure} 
    have also used explicit casts for managed type-level computation.
However, casts in those approaches primarily address type-level computations
  within contexts such as dependent types or type-level programming, 
  rather than the operational interpretation of recursive type equalities.
When considering the dynamic semantics of cast-like operations,
  there have been two major approaches.
One approach is to use an elaboration semantics,
  used in works like \cite{sjoberg2012irrelevance,sjoberg2015programming,stump2009verified},
  where the semantics are only defined for a cast-free language
  and the casts need to be erased before execution.
Another approach is to use push rules as 
  seen in \cite{sulzmann2007system,yorgey2012giving,weirich2013system,weirich2017specification},
  which is the approach that we adopt in our work.
Pure Iso-type Systems (PITS)~\cite{yang2019pure}
  provides a generalization of iso-recursive types 
  with explicit casts, but their focus is on unifying the syntax of
  terms and types while retaining decidable type checking,
  instead of subsuming equi-recursive type casting as we do.
Also, the form of casts is different from ours.
For example, they do not have a fixpoint cast, to enable coinductive reasoning.

\section{Conclusion}

This paper proposes full iso-recursive types, a generalization of
  iso-recursive types that can be used to encode the full power of
  equi-recursive types.  The key idea is to introduce a
  computationally irrelevant cast operator in the term language that
  captures all the equi-recursive type equalities.  We present \name,
  a calculus that extends simply typed lambda calculus with full
  iso-recursive types.  \name is proved to be type sound and has the
  same expressive power as a calculus with equi-recursive types, in
  terms of typing and dynamic semantics.  Our results can also be
  extended to subtyping, by encoding equi-recursive subtyping using
  iso-recursive subtyping with cast operators.

As future work, we plan to extend \name with other programming
language features, such as polymorphism and intersection and union types.  It
is also interesting to see whether our results can scale to real
world languages (e.g. Haskell). In particular, it would be interesting
to employ full iso-recursive types in an internal target language with
explicit cast operators for a source language using equi-recursive types.

% \begin{acks}
  
% % To Robert, for the bagels and explaining CMYK and color spaces.
% \end{acks}

\bibliographystyle{ACM-Reference-Format}
\bibliography{base}

\appendix

\end{document}